\documentclass[a4paper,fleqn,usenatbib]{mnras} 

\usepackage{newtxtext,newtxmath}

\usepackage[T1]{fontenc}
\usepackage{ae,aecompl}

\usepackage{graphicx}	
\usepackage{amsmath}	
\usepackage{amssymb}	
\usepackage{color,soul}

\newcommand*{\Vect}[1]{\ensuremath{\boldsymbol{#1}}}
\newcommand*{\Grad}{\Vect{\nabla}}
\newcommand*{\Div}{\Vect{\nabla}\cdot}
\newcommand*{\Curl}{\Vect{\nabla}\!\times\!}
\newcommand*{\Sh}{\Delta^*}

\title[Two-fluid magnetic evolution in neutron star cores]{Two-fluid simulations of the magnetic field evolution in neutron star cores in the weak-coupling regime}

\author[Castillo et al.]{
F. Castillo,$^{1}$\thanks{E-mail: fcastill@astro.puc.cl}
A. Reisenegger,$^{2}$
and J. A. Valdivia$^{3,4}$
\\
$^{1}$Instituto de Astrof\'isica, Facultad de F{\'i}sica, Pontificia Universidad Cat\'olica de Chile, Av. Vicu\~na Mackenna 4860, Macul, Santiago, Chile\\
$^{2}$Departamento de F\'{\i}sica, Facultad de Ciencias B\'asicas, Universidad Metropolitana de Ciencias de la Educaci\'on, Av. Jos\'e Pedro Alessandri 774,\\ \~Nu\~noa, Santiago, Chile\\
$^{3}$Departamento de F\'{\i}sica, Facultad de Ciencias,
Universidad de Chile, Las Palmeras 3425, \~Nu\~noa, Santiago, Chile\\
$^{4}$Centro para el Desarrollo de la Nanociencia y
Nanotecnolog\'\i a, CEDENNA, Santiago, Chile
}

\date{Accepted XXX. Received YYY; in original form ZZZ}

\pubyear{2020}

\begin{document}
\label{firstpage}
\pagerange{\pageref{firstpage}--\pageref{lastpage}}
\maketitle

\begin{abstract}
In a previous paper, we reported simulations of the evolution of the magnetic field in neutron star cores through ambipolar diffusion, taking the neutrons as a motionless uniform background.
However, in real neutron stars, neutrons are free to move, and a strong composition gradient leads to stable stratification (stability against convective motions) both of which might impact on the time-scales of evolution.
Here we address these issues by providing the first long-term two-fluid simulations of the evolution of an axially symmetric magnetic field in a neutron star core composed of neutrons, protons, and electrons with density and composition gradients.
Again, we find that the magnetic field evolves towards barotropic ``Grad-Shafranov equilibria'', in which the magnetic force is balanced by the degeneracy pressure gradient and gravitational force of the charged particles.
However, the evolution is found to be faster than in the case of motionless neutrons, as the movement of charged particles (which are coupled to the magnetic field, but are also limited by the collisional drag forces exerted by neutrons) is less constrained, since neutrons are now allowed to move.
The possible impact of non-axisymmetric instabilities on these equilibria, as well as beta decays, proton superconductivity, and neutron superfluidity, are left for future work.
\end{abstract}




\section{Introduction}

Neutron stars (NSs) are classified in several types based on their substantially different observational properties, which appear to be determined mainly by their rotation rates, magnetic field strengths, and presence or absence of accretion from a binary companion.
In particular, the magnetic fields control the torque that slows down their rotation, can have a crucial effect on the free energy budget and radiation processes, and influence the flow of accreted matter on and around these stars.
On the other hand, the structure and evolution of these magnetic fields is far from understood as different classes of NSs have very different magnetic field strengths, which appear to be correlated with their age.
Young NSs; such as magnetars, classical radio pulsars, and high-mass X-ray binaries; have strong surface magnetic fields $\sim 10^{11-15}$G; while the much older low-mass X-ray binaries and millisecond pulsars have weaker fields $\sim 10^{8-9}$G \citep{Kaspi2010,Vigano2013}.
This suggests a magnetic field decay, which is usually attributed to the accretion process \citep{Backer1982,Bhattacharya1991,Bhattacharya1995,Tauris2006}, but might also be explainable through its own dynamics and spontaneous decay as the NS cools and ages \citep{Cruces2019}.
Furthermore, the high time-averaged luminosity of magnetars is thought to be linked to the decay of their magnetic field, since their rotational energy loss is insufficient to account for it.
Since these objects appear to be isolated, their field decay must be attributed to processes intrinsic to the NSs \citep{Thompson1995,Thompson1996}.
Therefore, understanding the mechanisms that drive the long-term evolution of the magnetic field in NSs may help us unveil the relation between their different classes.

The physics of the evolution of the magnetic field varies in different regions of the NS \citep{Goldreich1992a}.
For example, in the solid crust, ions have very restricted mobility, so the currents are carried by electrons.
Therefore the long-term mechanisms that control the evolution of the field in this region are Ohmic diffusion, i.~e. current dissipation by electric resistivity; and Hall drift which corresponds to the transport of the magnetic flux by the electron motion.
Since the magnetic field dynamics is intrinsically non-linear, its recent studies have mainly taken a numerical point of view
\citep{Hollerbach2002,Pons2007,Pons2009,Vigano2012,Gourgouliatos2013,Vigano2013,Lander2019}, finding that the magnetic field evolves towards stable ``attractor'' configurations \citep{Gourgouliatos2014,Marchant2014}.

In this work, we concentrate on the long-term magnetic field dynamics in the core of NSs, which is conceptually much more involved than in the crust.
Indeed, the core of a NS is a fluid mixture of neutrons, protons, and electrons, joined by other species at increasing densities.
The mechanisms for field evolution in this region are strongly dependent on the star's core temperature $T$. Shortly after the NS is formed, neutrons and charged particles are strongly coupled by collisions, behaving essentially as a single fluid, coupled to the magnetic flux.
This fluid is stably stratified by its composition gradient (due to the different density profiles of charged particles and neutrons) \citep{Pethick1992,Goldreich1992a,Reisenegger2009}; thus, strong buoyancy forces oppose convective motions.
However, as noted by \citet{Reisenegger2007}, weak interaction processes (so-called ``Urca reactions''; \citealt{Gamow1941,Haensel1995}) can quickly adjust the composition of a fluid element, overcoming stable stratification and allowing the fluid to transport the magnetic flux.
In this ``strong-coupling'' regime, the matter can be regarded as a single, stably stratified, non-barotropic fluid (i.e., pressure depends on the non-uniform chemical composition in addition to density).
However, as weak interactions are strongly temperature-dependent, this regime applies only for a very short time after the NS birth (while $T\gtrsim 5\times 10^8$K).
As the star cools, the progressive reduction of the collisional coupling makes relative motions possible, so the charged particles can carry the magnetic flux with a different velocity field than that of the neutrons, a process called ``ambipolar diffusion'' \citep{Pethick1992,Goldreich1992a}.
Hence, on long time-scales, the NS core is in a ``weak-coupling'' regime, in which neutrons and charged particles act as independent barotropic fluids (assuming charged particles other than $p$ and $e$ can be ignored), with only the charged particles coupling significantly to the magnetic field. It has been argued that ambipolar diffusion may be relevant to explain the activity of magnetars due to its strong dependence on the magnetic field intensity \citep{Thompson1995, Thompson1996}.

In \citet{Castillo2017} (hereafter paper I), we provided the first simulations that evolve simultaneously and consistently the structure of the magnetic field and the small density perturbations it induces on the charged particles (taken to be a nearly uniform, locally neutral mix of protons and electrons) inside the core of an isolated, spherical NS through ambipolar diffusion in axial symmetry.
We modeled the neutrons as a motionless uniform background that produces a frictional drag on the charged particles (see \citealt{Goldreich1992a}) and neglected the currents in the crust, assuming it has a very low conductivity.
We also ignored the effects of superfluidity, superconductivity, and weak interactions (``Urca reactions'').

We found that the magnetic field evolves in the expected time-scales, eventually reaching equilibrium states in which
the magnetic forces are balanced by the degeneracy pressure gradient and gravitational force of charged particles.
The latter, being a homogeneous mixture of protons and electrons, can be regarded as a barotropic fluid (with a one-to-one relation between pressure and density), so these equilibrium states satisfy the Grad-Shafranov equation \citep{Grad1958,Shafranov1966}, strongly constraining the magnetic field configuration.
The toroidal field is confined to regions of closed poloidal field lines forming ``twisted tori'' whose number is conserved by the evolution, since ambipolar diffusion by itself cannot produce magnetic reconnection.
Outside these tori, i.e. along poloidal field lines that extend beyond the star, the toroidal field disappears through ``unwinding'' of the field lines by a toroidal component of the ambipolar velocity.
We also confirmed that previously found solutions of the Grad-Shafranov equation (e.g., \citealt{Armaza2015}) are stable in axial symmetry.

We must remark that barotropic axially symmetric equilibria (as the ones found in paper I, and eventually in this work) are likely to become unstable under non-axisymmetric perturbations, as there are numerous magneto-hydrodynamic instabilities identified for axially symmetric magnetic fields, all of which break the axial symmetry \citep{Tayler1973, Markey1973,Wright1973}. Additionally, stable stratification (non-barotropy) appears to be required to stabilize magnetic field configurations \citep{Braithwaite2009,Reisenegger2009,Mitchell2015}.
Such instabilities might lead to a complete dissipation of the field, unless the charged particle fluid is stably stratified by a gradient of muons or other particle species, in addition to protons and electrons.
Addressing 3-dimensional motions, as well as the high temperature ``strong-coupling'' regime (in which collisional coupling and weak interactions between species become important) is left for future work. 

Paper I was a starting point, but contains several simplifications that must be considered.
For instance, we modeled the neutrons as a motionless, uniform background with the only effect of providing a frictional force on the charged particles.
However, in real NSs, neutrons can move, and both neutrons and charged particles have strong, but different density gradients, so their velocity fields must also be different, as long as their relative abundances cannot be adjusted ``in real time'' by Urca reactions, as is the case at high temperatures.
Thus, there will both be a bulk motion and a relative motion of neutrons and charged particles, setting time-scales that might be somewhat shorter than those previously found.
This has been recently studied by \citet{Ofengeim2018}, who evaluated the different instantaneous particle velocities for a fixed magnetic field configuration, finding that the magnetic field may lead to the generation of a macroscopic fluid velocity that is substantially larger than the relative velocities between species.
The impact of the neutron motion on an evolving magnetic field has been previously studied by \citet{Hoyos2008,Hoyos2010}; however, their one-dimensional simulations cannot capture all the relevant physics of the process.
Therefore, in the present work we address those issues by numerically evolving the magnetic field, including the motion of neutrons, in a spherical star with an axially symmetric magnetic field. 

In this paper, we neglect the effects of superfluidity and superconductivity, which should at the very least change the couplings between superfluid neutrons and superconducting protons, thus modifying the time-scales for the processes simulated here, and possibly change the equilibrium configurations found.
These issues have been studied by several authors in the last years (e.g., \citealt{Glampedakis2011,Graber2015,Elfritz2016, Passamonti2017, Bransgrove2018}), and the discussion has not been exempt from controversy \citep{Gusakov2016, Dommes2017, Passamonti2017b}.
Until recently, the impact of superfluidity and superconductivity was thought to be well understood only in particularly simple situations, such as the one studied by \citet{Kantor2017}.
However, it was not until very recently that an expression for the force on proton vortices in superfluid and superconducting matter \citep{Gusakov2019} was proposed, which is a key ingredient for studying the long-term magnetic evolution.
Thus, performing magneto-hydrodynamic simulations including such effects is left for future work.

This paper is organized as follows.
In \S~\ref{sec:model} we discuss the physical model and construct the equations to be solved in axial symmetry. Here, we also discuss the relevant time-scales and describe our toy model for the equation of state.
In \S~\ref{sec:results:z_comp} we test the impact of including an extra artificial friction, which is used in our simulations to keep the magnetic field close to a magneto-hydrostatic quasi-equilibrium throughout its evolution.
In \S~\ref{sec:results:evolution} we check if the simulations are in agreement with the analytically expected time-scales. 
In \S~\ref{sec:results:GS} we study the equilibrium configurations found.
Finally, in \S~\ref{sec:conclusions}, our results are summarized and conclusions are outlined.

\section{Physical model}
\label{sec:model}

\subsection{Basic equations}
\label{sec:model:basic}

As in Paper I, we model the interior of an isolated NS as a plasma composed of neutrons, protons, and electrons.
The species are coupled by collisions and electromagnetic forces, and their equations of motion are written as
\begin{gather}
\begin{split}
n_i\frac{\mu_i}{c^2}\frac{d\Vect{v_i}}{dt}=\,& n_i q_i\left(\Vect E +\frac{\Vect{v_i}}{c}\times\Vect B \right) -n_i\Grad\mu_i - \frac{n_i\mu_i}{c^2}\Grad\Psi 
\\&- \sum_{j\neq i}\gamma_{ij}n_i n_j \left( \Vect {v_i} - \Vect {v_j}\right) \,,\label{eq:eqmovimiento}
\end{split}
\end{gather}
where $n_i$ and $\mu_i$ ($i=n,p,e$) are the number density and chemical potential of species $i$, respectively; $\mu_i/c^2$ is the effective mass of each species, which could include corrections due to interactions and relativistic effects \citep{Akmal1998}; $q_i$ is the electric charge of the respective particles; 
and $\Vect{v_i}$ is the velocity of the $i^{th}$ species.
We assume charge neutrality, so that at all times $n_p=n_e\equiv n_c$.
The forces acting on each particle are, from left to right, the Lorentz force (where $\Vect E$ and $\Vect B$ are the electric and magnetic fields), the degeneracy-pressure gradient of species $i$, the gravitational force acting on each fluid species (where $\Psi$ is the gravitational potential), and the frictional drag forces due to collisions between particles of different species. The later are parametrized by rate coefficients $\gamma_{ij}$, so that momentum conservation implies $\gamma_{ij}=\gamma_{ji}$.

Under any perturbation, the star very quickly reaches a magneto-hydrostatic quasi-equilibrium state in which all the forces on a fluid element are close to balancing each other. The time-scale to reach this state is a few Alfv\'en times, $t_{\rm{Alf}}\sim(10^{14}\mathrm{G}/B)$~s.
Since we are interested in the long-term evolution of the field, which happens on much longer time-scales, $\sim 10^{3-10}$~yr, we do not intend to follow the propagation of sound waves, gravity (buoyancy) waves, and Alfv\'en waves in detail.
Instead, we filter them out by replacing the inertial terms on the left-hand-side of the equations of motion by an artificial frictional force acting on the neutrons, of the form $\Vect{f_\zeta}\equiv-\zeta n_n \Vect{v_n}$ \citep{Hoyos2008}.
The balance between this and the other forces acting on a fluid element determines the velocity field $\Vect{v_n}$, which quickly restores the hydro-magnetic quasi-equilibrium by rearranging the particles and magnetic field on a time-scale set by the parameter $\zeta$.
The value of this parameter is chosen to be small enough so its associated time-scale is much longer than the dynamical time-scales ($\sim$ Alfv\'en times), but shorter than the time-scales relevant to us (see discussion in Section \ref{sec:model:timescales}).

In paper I, we did not need to explicitly introduce such mechanism. As we froze the neutrons, their friction with the charged particles provided a mechanism ensuring that the latter would evolve through successive quasi-equilibrium states.
In contrast, in the present work we are allowing neutrons to move and rearrange in a way in which they help to maintain a hydro-magnetic quasi-equilibrium, which might be much closer to reality.

The equations of motion then become:
\begin{gather}
\begin{split}
0=&-n_n\Grad\mu_n - \frac{n_n\mu_n}{c^2}\Grad\Psi - \gamma_{ne}n_n n_c \left( \Vect{v_n} - \Vect{v_e}\right)\\& - \gamma_{np}n_n n_c \left( \Vect{v_n} - \Vect{v_p}\right) - \zeta n_n\Vect{v_n} \,,\label{eq:neutrons}
\end{split}\\
\begin{split}
0=&+ n_c e\left(\Vect E +\frac{\Vect{v_p}}{c}\times\Vect B \right)-n_c\Grad\mu_p - \frac{n_c\mu_p}{c^2}\Grad\Psi\\ &-\gamma_{pe}n_c^2 \left( \Vect{v_p} - \Vect{v_e}\right) - \gamma_{pn}n_c n_n \left( \Vect{v_p} - \Vect{v_n}\right) \,,\label{eq:protons}
\end{split}\\
\begin{split}
0=&- n_c e\left(\Vect E +\frac{\Vect{v_e}}{c}\times\Vect B \right)-n_c\Grad\mu_e - \frac{n_c\mu_e}{c^2}\Grad\Psi  \\
&- \gamma_{ep}n_c^2 \left( \Vect{v_e} - \Vect{v_p}\right) - \gamma_{en}n_c n_n \left( \Vect{v_e} - \Vect{v_n}\right) \,.\label{eq:electrons}
\end{split}
\end{gather}
By taking the product of equation~\eqref{eq:protons} with $\gamma_{en}$ and equation~\eqref{eq:electrons} with $\gamma_{pn}$, and then subtracting them, we get
\begin{equation}
\begin{split}
\Vect E=& - \frac{\gamma_{pn}\Vect{v_e} +\gamma_{en}\Vect{v_p}}{c\gamma_{cn}}
\times\Vect B +\frac{\Vect{J}}{\sigma}
\\&+\frac{\gamma_{en}\Grad\mu_p-\gamma_{pn}\Grad\mu_e}{e\gamma_{cn}}
+  \frac{\gamma_{en}\mu_p-\gamma_{pn}\mu_e}{ec^2\gamma_{cn}}
\Grad\Psi.
\end{split}
\end{equation}
where $\gamma_{cn}=\gamma_{pn}+\gamma_{en}$ is the net collisional coupling between charged particles and neutrons,
\begin{equation}
\sigma = e^2\left(\gamma_{pe}+\frac{\gamma_{en}\gamma_{pn}}{\gamma_{cn}}\frac{n_n}{n_c} \right)^{-1} \,,
\end{equation}
is the electric conductivity, and $\Vect J$ is the electric current density $\Vect{J}=n_ce(\Vect{v_p}-\Vect{v_e})=c\Curl\Vect{B}/4\pi$.

We define the ``ambipolar diffusion velocity'', which represents the joint motion of the two charged particle species relative to the neutrons, as
\begin{equation}
\Vect{v_{ad}} = \frac{\gamma_{pn}(\Vect{v_p} -\Vect{v_n})+\gamma_{en}(\Vect{v_e} -\Vect{v_n})}{\gamma_{cn}} \,,\label{eq:def_va}
\end{equation}
and the ``Hall drift velocity'', which is proportional to the electric current, as
\begin{equation}
\Vect{v_H} = - \frac{\gamma_{pn}-\gamma_{en}}{\gamma_{cn}} (\Vect{v_p} -\Vect{v_e})
= - \frac{\gamma_{pn}-\gamma_{en}}{\gamma_{cn}}\frac{\Vect J}{n_c e}  \,.
\end{equation}
Hence, $\left(\gamma_{pn}\Vect{v_e}+\gamma_{en}\Vect{v_p}\right)/\gamma_{cn}=\Vect{v_n}+\Vect{v_{ad}}+\Vect{v_H}$, so the evolution equation for the magnetic field is obtained from Faraday's law as
\begin{gather}
\begin{split}
\frac{\partial\Vect B}{\partial t}
=\:& \Curl\left[\left( \Vect{v_n}+\Vect{v_{ad}}+\Vect{v_H}\right)\times\Vect B -\frac{c}{\sigma}\Vect{J}\right]\\
&-\Grad\left(\frac{c\gamma_{en}}{e\gamma_{cn}}\right) \times\Grad\mu_c -\Grad\left( \frac{\mu_p\gamma_{en}-\mu_e\gamma_{pn}}{ec\gamma_{cn}}\right) \times\Grad\Psi
\,,\end{split}\label{eq:faraday_full}
\end{gather}
where we defined a total chemical potential for the charged particles, $\mu_c\equiv\mu_p+\mu_e$. The last term inside the curl represents Ohmic dissipation, and the two last terms represent battery effects.

As discussed in \citet{Goldreich1992a}, in the core of NSs the effects of Hall drift and Ohmic decay can be orders of magnitude smaller than the ambipolar diffusion, hence we neglect those terms.
Also, as shown in paper I, the time-scale on which the battery terms are relevant is roughly the same as the Hall time-scale, therefore we also neglect those terms, obtaining
 \begin{equation}
 \frac{\partial\Vect B}{\partial t}= \Curl\left(\Vect{v_c}\times\Vect B \right) \,,
 \end{equation}
where $\Vect{v_c}\equiv\Vect{v_n}+\Vect{v_{ad}}$ is the velocity of the charged particles.
 
The relevant velocities can be computed from the equations of motion. By adding equations \eqref{eq:neutrons}, \eqref{eq:protons}, and \eqref{eq:electrons}, we get the velocity field of the neutrons, parametrized by the fictitious friction coefficient $\zeta$, which replaces the very small inertial terms,
\begin{equation}
\Vect{v_n}=\frac{1}{\zeta n_n}\left[\frac{\Vect J}{c}\times\Vect B -n_c\left( \Grad\mu_c +\frac{\mu_c}{c^2}\Grad\Psi\right)  -n_n\left( \Grad\mu_n +\frac{\mu_n}{c^2}\Grad\Psi\right) \right]\label{eq:neutrones0}\,.
\end{equation}
From equations \eqref{eq:protons}, \eqref{eq:electrons}, and \eqref{eq:def_va} we obtain the ambipolar diffusion velocity,
\begin{equation}
\Vect{v_{ad}} = \frac{1}{\gamma_{cn}n_c n_n}\left[\frac{\Vect{J}\times\Vect{B}}{c} - n_c\left(\Grad\mu_c+\mu_c\frac{\Grad\Psi}{c^2}\right)\right]\label{eq:ambipolar0}\,,
\end{equation}
which is proportional to the imbalance between the forces, including (from left to right) the magnetic force density, the gradient of the degeneracy pressure of the charged particles, and the gravitational force on the charged particles.
Thus, the ambipolar diffusion is driven by the magnetic force, controlled by the pressure gradient and gravitational forces acting on the charged particles, and opposed by the collisional drag of the neutrons.
The terms in equation~\eqref{eq:neutrones0} have an analogous interpretation.

To evolve the particle densities, we use the continuity equations 
\begin{gather}
	\frac{\partial n_c}{\partial t} + \Div\left(n_c\Vect{v_c}\right) = -\Delta\Gamma \label{eq:continuidad}\,,\\
	\frac{\partial n_n}{\partial t} + \Div\left(n_n\Vect{v_n}\right) =   +\Delta\Gamma \,,
\end{gather}
where $\Delta\Gamma$ is the net rate per unit volume of conversion of charged particles to neutrons by weak interactions, i.~e., the difference between the rates for the (direct or modified) Urca processes, $p+e\to n+\nu_e$ and $n\to p+e+\bar\nu_e$, where $\nu_e$ and $\bar\nu_e$ denote electron neutrinos and electron antineutrinos, respectively.
This quantity becomes very small in the low-temperature regime ($T\lesssim 5\times 10^8$K) of interest here \citep{Goldreich1992a,Reisenegger1995}, therefore we ignore it in the present work.

\subsection{Background NS model}
\label{sec:model:stellar}

\begin{figure}
	\centering
	\includegraphics[width=\linewidth]{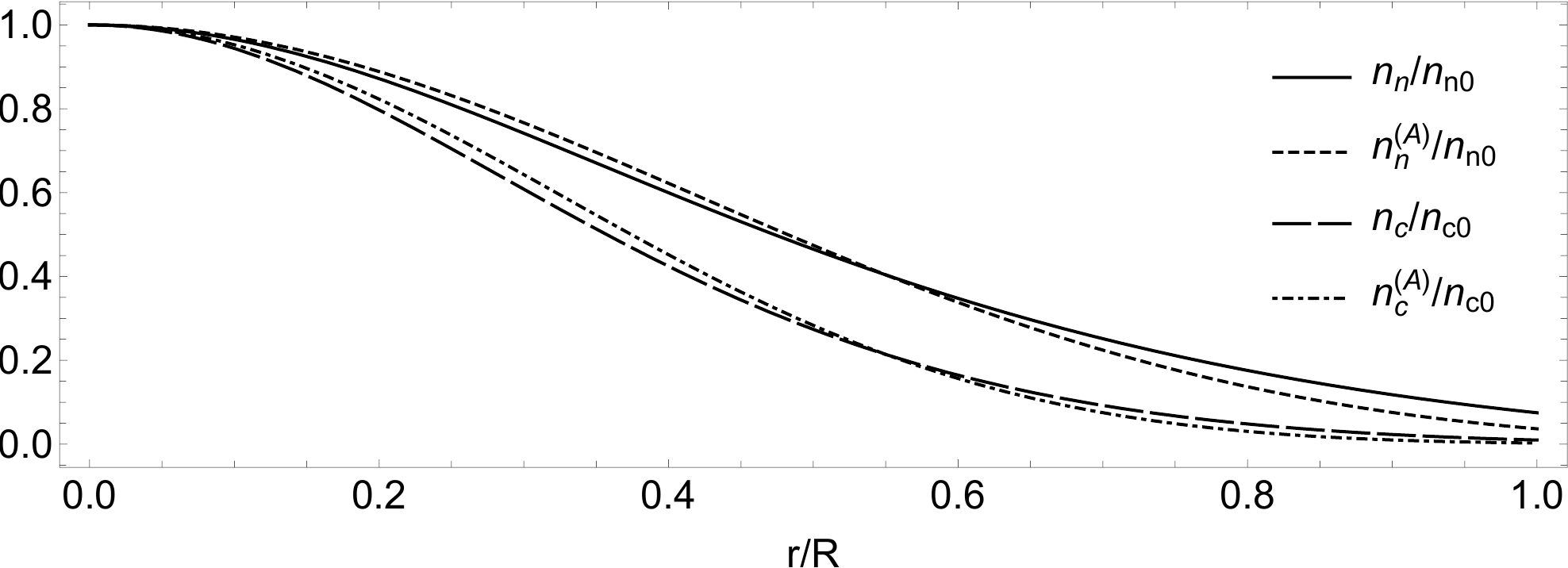}
	\caption{Comparison between $n_n/n_n(0)$ and $n_c/n_c(0)$ (obtained by numerically solving the hydrostatic equilibrium equation~\ref{eq:hydrostatic-equilibrium}), and the simple analytical approximations $n_n^{(A)}/n_n(0)$ and $n_c^{(A)}/n_c(0)$, respectively.
	Here, $n_n^{(A)}$ is computed from equation~\eqref{eq:chemical-equilibrium-n}, using $n_c^{(A)}$, the analytical expression for $n_c$ given by equation~\eqref{eq:nc_A}.
	}
	\label{fig:n-fit}
\end{figure}

As in paper I, we consider the (realistic) situation where the magnetic field induces only small perturbations with respect to a hypothetical non-magnetized stellar structure (see also \citealt{Reisenegger2009}). 
Thus, we split the particle densities, and hence the chemical potentials, in two:  
\begin{itemize}
\item[1)] time-independent background densities $n_i(r)$ and chemical potentials $\mu_i(r)$ determined by the conditions of chemical (beta) equilibrium,
\begin{equation} \label{eq:chemical-equilibrium}
\mu_{n}=\mu_{c}\equiv\mu(r),    
\end{equation}
and hydrostatic equilibrium in the absence of the magnetic field, 
\begin{equation} \label{eq:hydrostatic-equilibrium}
\Grad\mu+\mu\Grad\Psi/c^2=0,    
\end{equation}
and 
\item[2)] much smaller time-dependent perturbations $\delta n_c$, $\delta n_n$, $\delta \mu_c$, and $\delta \mu_n$, respectively, induced by the evolving magnetic field (to be discussed in \S~\ref{sec:model:linearization}).
\end{itemize}
An important improvement of the present work over paper I is that we consider the non-magnetized background star to have non-uniform particle densities, with different radial gradients for the neutrons and charged particles, as imposed by beta equilibrium. 

For simplicity, we ignore strong interactions, treating neutrons and protons (with the same mass $m$) as non-relativistic Fermi gases and electrons as an extremely relativistic (massless) Fermi gas.
Thus, the chemical potentials are related to the particle densities by 
\begin{gather}
\mu_n(r) = m c^2 + \frac{p_{Fn}(r)^2}{2m}\,,\\
\mu_c(r) = m c^2 + \frac{p_{Fc}(r)^2}{2m} + p_{Fc}(r)c \,,
\end{gather}
where $p_{Fi}(r) = \hbar\left[3\pi^2n_i(r)\right]^{1/3}\,$ are the Fermi momenta of neutrons ($i=n$) and charged particles ($i=c$).
Since we assume charge neutrality, the densities (and thus the Fermi momenta) of protons and electrons are the same. Thus, the condition of chemical equilibrium (equation~[\ref{eq:chemical-equilibrium}]) allows to write the density of neutrons in terms of that of the charged particles,
\begin{equation}
n_n(r)=\left[\frac{2mc}{\hbar(3\pi^2)^{1/3}}n_c(r)^{1/3} + n_c(r)^{2/3}\right]^{3/2}\,.\label{eq:chemical-equilibrium-n}
\end{equation}

For the background number density of charged particles, we use the simple analytical approximation
\begin{equation}
n_c(r)=n_{c0}\left[\frac{\sin(0.7\pi r/R)}{0.7\pi r/R}\right]^6\,,\label{eq:nc_A}
\end{equation}
where $R$ is the radius of the core, taken as $70\%$ of the stellar radius. As can be seen in Fig.~\ref{fig:n-fit}, this relation closely resembles the numerical solution obtained by imposing Newtonian hydrostatic equilibrium equation~(\ref{eq:hydrostatic-equilibrium}) on this Fermi gas, and we adjusted the central density so $\xi\equiv n_n(0)/n_c(0)=10$, yielding a star of mass $1.58 M_{\odot}$ and radius $8.2$ km.

The transport coefficients $\gamma_{ij}(r,T)$ between species $i$ and $j$ are computed for the relations derived for $\tau_{ij}(r,T)$ by \citet{Yakovlev1990}, where $T$ denotes the core temperature.

This stellar model, although very simplified, allows us to capture the effects of radial density gradients, gravity, and stable stratification into our simulations, which is a vast improvement over paper I, where the background densities were treated as fixed radially independent numbers.

\subsection{Linearization}
\label{sec:model:linearization}

Since the density perturbations are small, we can linearize $\delta\mu_n = K_{nn}\delta n_n$ (where $K_{nn}=d\mu_n/dn_n$), and $\delta\mu_c=K_{cc}\delta n_c$ (where $K_{cc}=d\mu_c/dn_c$). In our model, as there are no strong interactions, the cross-derivatives $K_{cn}=\partial\mu_c/\partial n_n$ and $K_{nc}=\partial\mu_n/\partial n_c$ vanish.

Dropping higher-order terms, we can write the full set of equations to be solved as
\begin{gather}
\frac{\partial\Vect B}{\partial t}= \Curl\left[\left( \Vect{v_n}+\Vect{v_{ad}}\right)\times\Vect B \right] \label{eq:faraday}\,,\\
\frac{\partial\delta n_n}{\partial t} + \Div\left(n_n\Vect{v_n}\right) = 0\label{eq:continuity_n}\,,\\
\frac{\partial\delta n_c}{\partial t} + \Div\left[n_c(\Vect{v_n}+\Vect{v_{ad}})\right] = 0 \,, \label{eq:continuity_c}\\
\Vect{v_n}=\frac{1}{\zeta n_n}\left(\Vect{f_B}+\Vect{f_n}+\Vect{f_c}\right) \,,\label{eq:v_neutrons}\\
\Vect{v_{ad}} = \frac{1}{\gamma_{cn}n_c n_n}\left(\Vect{f_B}+\Vect{f_c}\right) \label{eq:v_ambipolar}\,,\\
\Vect{f_B}=\frac{(\Curl\Vect B)\times\Vect B}{4\pi} \label{eq:Lorentz force} \,,\\
\Vect{f_n}=-n_n\mu\Grad \left(\frac{\delta\mu_n}{\mu}\right)\label{eq:neutron force} \,,\\
\Vect{f_c}=-n_c\mu\Grad \left(\frac{\delta\mu_c}{\mu}\right)\label{eq:charged force} \,,\\
\delta\mu_n = K_{nn}\delta n_n \label{eq:dmun}\,,\\
\delta\mu_c = K_{cc}\delta n_c \label{eq:dmuc}\,,
\end{gather}
where the degeneracy pressure gradients and the gravitational forces for each species have been combined into the ``fluid forces'' (more precisely, force densities) $\Vect{f_n}$ and $\Vect{f_c}$ using the hydrostatic equilibrium condition of the background, equation~(\ref{eq:hydrostatic-equilibrium}).

\subsection{Boundary conditions}
\label{sec:model:boundary}

For simplicity, we assume that currents in the crust decay much faster than typical evolution time-scales in the core, so we can treat the crust as a vacuum whose magnetic field at any time is fully determined by the field in the core (see discussion and more detail on the imposed boundary conditions in paper I).
This external current-free magnetic field is computed at all time-steps as a multipolar expansion, whose coefficients ($a_\ell$, with $\ell=1,2,...$) are determined by the value of the radial component of the magnetic field at the crust-core interface.
These coefficients also determine the energy stored in the external magnetic field (see details in paper I), namely  
\begin{equation}
U_{\text{ext}}\equiv \sum_{\ell=1}^\infty U_{\text{ext},\ell}=\sum_{\ell=1}^\infty \frac{\ell+1}{2\ell+1}\frac{a_\ell^2}{2}\,, \label{eq:uext}
\end{equation}
where $U_{\text{ext},\ell}$ is the energy stored in the $\ell$-th component of the external field.
Also, at the crust-core interface we assume that the radial components of both the neutron velocity $\Vect{v}_n$ and the charged-particle velocity $\Vect{v}_c$ are null.
Therefore, at the crust-core interface we have
\begin{gather}
\frac{\partial}{\partial r}\left(\frac{\delta\mu_c}{\mu}\right) =\frac{f_B^r}{n_c\mu}\,,\\
\frac{\partial}{\partial r}\left(\frac{\delta\mu_n}{\mu}\right)=0 \,.
\end{gather}

\subsection{Axial symmetry}
\label{sec:model:axial}

Lastly, we restrict ourselves to axial symmetry, so the magnetic field can be decomposed as
\begin{equation}
\Vect{B}=\Grad\alpha\times\Grad\phi + \beta\Grad\phi \,. \label{eq:BAlfaBeta}
\end{equation}
Here the scalar potentials $\alpha(t,r,\theta)$ and $\beta(t,r,\theta)$ generate the poloidal and toroidal magnetic field, respectively, where $t$ denotes time, $r$ is the radial coordinate, and $\theta$ and $\phi$ are the polar and azimuthal angles, respectively, so $\Grad\phi=\hat{\phi}/(r\sin\theta)$.
An explicit form for the evolution of the magnetic potentials can be derived from equation~\eqref{eq:faraday}, where we get
\begin{gather}
\frac{\partial\alpha}{\partial t}=r\sin\theta\left[\left(\Vect{v_n}+\Vect{v_{ad}}\right)\times\Vect{B}\right]\cdot\hat\phi\,, \label{eq:evolucionalfa}\\
\frac{\partial\beta}{\partial t}=r^2\sin^2\theta\Div\left\lbrace\frac{\left[\left(\Vect{v_n}+\Vect{v_{ad}}\right)\times\Vect{B} \right]\times\hat{\phi}}{r\sin\theta} \right\rbrace\,. \label{eq:evolucionbeta}
\end{gather}

\subsection{Time-scale estimates}
\label{sec:model:timescales}

\subsubsection{Short-term relaxation through fictitious friction}
\label{sec:model:timescales:1}

In order to mimic the quick hydro-magnetic relaxation due to the propagation and damping of sound waves, gravity waves, and Alfv\'en waves, which happens much faster than ambipolar diffusion, the fictitious friction parameter introduced in equation~(\ref{eq:neutrons}) must satisfy $\zeta\ll\gamma_{cn}n_c$, so a net force imbalance on a fluid element (in round brackets on the right-hand side of equation~[\ref{eq:v_neutrons}]) is reduced by bulk motions (with velocity $\Vect{v}_n$) much more quickly than an imbalance of the partial forces on the charged-particle component (in round brackets on the right-hand side of equation~[\ref{eq:v_ambipolar}]) is reduced by ambipolar diffusion (with relative velocity $\Vect{v}_{ad}$).
Thus, for an arbitrary, non-equilibrium initial condition, the dynamics is initially dominated by a bulk motion with the neutron velocity, $\Vect{v}_n$.
For a characteristic spatial scale $L\lesssim R$ of the magnetic field, the time-scales for the evolution of the density perturbations, $\sim (\delta n_n/n_n)(L/v_n)$ (from equation~[\ref{eq:continuity_n}]) and $\sim (\delta n_c/n_c)(L/v_n)$ (from equation~[\ref{eq:continuity_c}]) are much shorter than that for the magnetic field, $\sim L/v_n$ (from equation~[\ref{eq:faraday}]).
Thus, for $t\ll L/v_n$, the magnetic field can be regarded as fixed, and only the density perturbations evolve.\footnote{Formally, equations (\ref{eq:continuity_n}) to (\ref{eq:dmuc}) can be rewritten as a system of two coupled diffusion equations for $\delta n_n$ and $\delta n_c$ with an inhomogeneous term linear in $\Vect{f_B}$ (which can be taken as a constant in this regime) and diffusion coefficients (and thus characteristic time-scales) independent of $\Vect B$.}

For an axially symmetric magnetic field configuration, the Lorentz force density (equation~[\ref{eq:Lorentz force}]) can be decomposed as
\begin{equation}
\Vect{f}_B=\Vect{f}_B^{\mathrm{Pol}}+\Vect{f}_B^{\mathrm{Tor}},   
\end{equation}
where $\Vect{f}_B^{\mathrm{Pol}}$ and $\Vect{f}_B^{\mathrm{Tor}}$ are its poloidal and a toroidal components.
On the other hand, the fluid forces given by equations~(\ref{eq:neutron force}) and (\ref{eq:charged force}) are purely poloidal, therefore they cannot balance $\Vect{f}_B^{\mathrm{Tor}}$.
Furthermore, each of the fluid forces is proportional to the gradient of a single scalar function, therefore an arbitrary $\Vect{f}_B^{\mathrm{Pol}}$ can only be balanced for a particular, non-trivial combination of chemical potential perturbations, $\delta\mu_n(r,\theta)$ and $\delta\mu_c(r,\theta)$.

For definiteness, let us assume an initial condition with vanishing density perturbations, $\delta n_n(t=0)=\delta n_c(t=0)=0$.
In this case, the initial poloidal bulk velocity is $\Vect{v}_n^{\mathrm{Pol}}(t=0)=\Vect{f}_B^{\mathrm{Pol}}/(\zeta n_n)$, and the density perturbations grow roughly as $|\delta n_n/n_n|\sim|\delta n_c/n_c|\sim v_n^{\mathrm{Pol}} t/L$. This causes a growth of the fluid forces, $f_n\sim K_{nn}n_n^2v_n^{\mathrm{Pol}} t/L^2$ and $f_c\sim K_{cc}n_c^2v_n^{\mathrm{Pol}} t/L^2$, until the larger of these, namely $\Vect{f}_n$ (because $K_{nn}n_n\sim K_{cc}n_c$, whereas $n_c\ll n_n$) approaches the magnitude of $\Vect{f}_B^{\mathrm{Pol}}$.
This happens on a time-scale
\begin{equation} \label{eq:t_zeta_p}
t_{\zeta p}\sim\frac{\zeta L^2}{K_{nn}n_n},
\end{equation}
the analog of the propagation time of sound waves (p-modes) when inertial effects are taken into account.
However, as discussed in the previous paragraph, $\Vect{f}_n$ alone cannot balance an arbitrary (poloidal) vector field $\Vect{f}_B^{\mathrm{Pol}}(r,\theta)$.
Thus, $\Vect{v}_n$ might now be a very different vector field than $\Vect{f}_B/(\zeta n_n)$, but it will still be roughly of the same order of magnitude, further modifying the density perturbations until $\Vect{f}_B^{\mathrm{Pol}}+\Vect{f}_n+\Vect{f}_c\approx 0,$
at which point the density perturbations reach the (still very small) fractional magnitudes $|\delta n_n|/n_n\sim B^2/(4\pi K_{nn}n_n^2)$ and $|\delta n_c|/n_c\sim B^2/(4\pi K_{cc}n_c^2)$.
The latter happens on a time-scale 
\begin{equation} \label{eq:t_zeta_g}
t_{\zeta g}\sim\frac{L}{v_n}\frac{|\delta n_c|}{n_c}\sim\frac{\zeta n_n L^2}{K_{cc}n_c^2},
\end{equation}
which should be roughly identified with the propagation time of gravity waves (g-modes), i.~e., the buoyancy (Brunt-V\"ais\"al\"a) period in a realistic NS \citep{Reisenegger1992}, since balancing an arbitrary $\Vect{f}_B^{\mathrm{Pol}}$ will generally require non-parallel vector fields $\Grad[\delta\mu_n(r,\theta,t)/\mu(r)]$ and $\Grad[\delta\mu_c(r,\theta,t)/\mu(r)]$, i.~e., a baroclinic (non-barotropic) configuration.

On the other hand, $\Vect{f}_B^{\mathrm{Tor}}$ cannot be canceled by fluid forces and must thus decay to zero, which occurs on a longer time-scale
\begin{equation}\label{eq:t_zeta_B}
t_{\zeta B}\sim\frac{L}{v_n}\sim\frac{4\pi\zeta n_n L^2}{B^2},
\end{equation}
which should be identified with an Alfv\'en-like time. The latter evolution clearly also modifies $\Vect{f}_B^{\mathrm{Pol}}$, but the fluid displacements can keep up, always maintaining the balance between $\Vect{f}_B^{\mathrm{Pol}}$ and the gradient terms and eventually leading to a nearly complete cancellation of the forces on the right-hand side of equation~(\ref{eq:v_neutrons}), thus making $|\Vect{v}_n|$ comparable to $|\Vect{v}_{ad}|$. 

These arguments remain valid when removing the restriction of axial symmetry. As two gradient forces can only balance two different components of the magnetic force (in a time-scale $t_{\zeta g}$), one component will always remain unbalanced.
Therefore, reaching a state of hydro-magnetic quasi-equilibrium will always require the magnetic field to adjust so that the latter component vanishes, which will happen on a time-scale $\sim t_{\zeta B}$.
Since we are interested in an evolution on a time-scale many orders of magnitude longer than an Alfv\'en-like time, we only need $\zeta$ to be small enough to make $t_{\zeta B}$ sufficiently short to maintain the hydro-magnetic quasi-equilibrium discussed above.

\citet{Gusakov2017} used the time-derivative of the toroidal component of the quasi-equilibrium condition, $\partial\Vect{f}_B^{\mathrm{Tor}}/\partial t=0$, to obtain the macroscopic toroidal velocity of the particle flow.
However, solving this equation at all time-steps in a simulation is unpractical, as it would require high spatial resolution to properly resolve the fourth-order derivatives over $\Vect{B}$.
Our approach is simpler, as we do not attempt to solve this equation directly, rather, we let the system relax to this condition through the addition of the fictitious friction force.

\subsubsection{Long-term evolution through ambipolar diffusion}
\label{sec:model:timescales:2}

As the magnetic force acts only on the charged particles (which are coupled to the neutrons only by collisions), it induces a slow relative motion with velocity $\Vect{v}_{ad}$ in which the magnetic field lines are carried along with the charged particles. 
This process slowly modifies the hydro-magnetic quasi-equilibrium, eventually reaching a state in which the magnetic force is balanced mainly by the fluid force of the charged particles, while the contribution of the neutrons becomes negligible.
During this process, the magnetic field has to adjust, since it has to transition from a state in which $\Vect{f}_B^{\mathrm{Pol}}$ is balanced by the sum of two different gradients to a state in which only one gradient is available.
While the charged particles diffuse through neutrons at a velocity $v_{ad}\sim B^2/4\pi L \gamma_{cn}n_cn_n$, the neutrons have to adjust accordingly. 
Therefore, the velocity of the neutrons is also controlled by $\gamma_{cn}$ and independent of the value given to $\zeta$ (granted that $\zeta$ is sufficiently small, so $t_{\zeta B}$ is much shorter than the ambipolar diffusion time-scale).

At this stage, we expect $|\partial\delta n_c/\partial t|\sim |\delta n_c|v_{ad}/L\ll n_c v_{ad}/L\sim|\Div(n_c\Vect{v_{ad}})|$,
and therefore $\Div\left(n_c\Vect{v_{ad}}\right)\approx -\Div\left(n_c\Vect{v_n}\right)$.
This imposes a restriction on the irrotational component of  $n_c\Vect{v_{ad}}$ and $n_c\Vect{v_n}$, so we expect $\Vect{v_{ad}}$ and $\Vect{v_n}$ to be of similar order of magnitude.
The magnetic field will evolve under this process until the fluid force of the charged particles balances the magnetic force, reaching an equilibrium state (which may not be stable in the presence of not axially symmetric instabilities, and can also be very slowly eroded by Ohmic decay and other processes neglected in this analysis).
This dissipative process sets the long time-scale for magnetic field evolution, which can be estimated as
\begin{equation}
\begin{split}
t_{ad} &\sim \frac{L}{|\Vect{v_n}+\Vect{v_{ad}}|} \lesssim \frac{4\pi\gamma_{cn}n_cn_nL^2}{B^2} \\
&\sim 3\times 10^{3} \, \left(\frac{10^{15}\text{G}}{B}\right)^{2}\left(\frac{T}{10^8\text{K}}\right)^{2}\left(\frac{L}{1\text{km}}\right)^{2}\,\text{yr.}\,,
\end{split} \label{eq:t_ad}
\end{equation}  
where $v_n\sim v_{ad}$. (See \citealt{Ofengeim2018} for analytic relations between these two velocity fields.)

\subsection{Dimensionless equations}
\label{sec:model:dimensionless_equations}

We have written the equations in dimensionless form,
\begin{gather}
\frac{\partial\alpha}{\partial t}=r\sin\theta\left[\left(\Vect{v_n}+\Vect{v_{ad}}\right)\times\Vect{B}\right]\cdot\hat\phi\,, \label{eq:a:alpha}\\
\frac{\partial\beta}{\partial t}=r^2\sin^2\theta\Div\left\lbrace\frac{\left[\left(\Vect{v_n}+\Vect{v_{ad}}\right)\times\Vect{B} \right]\times\hat{\phi}}{r\sin\theta} \right\rbrace\,, \label{eq:a:beta}\\
\frac{\partial\delta n_n}{\partial t} + \Div\left(n_n\Vect{v_n}\right) = 0\label{eq:a:continuity_n}\,,\\
\frac{\partial\delta n_c}{\partial t} + \Div\left[n_c(\Vect{v_n}+\Vect{v_{ad}})\right] = 0 \,, \label{eq:a:continuity_c}\\
\Vect{v_n}=\frac{\xi}{\zeta n_n}\left[b^2\left(\Curl\Vect{B}\right)\times\Vect B -n_c\mu\Grad \left(\frac{\delta\mu_c}{\mu}\right) -n_n\mu\Grad \left(\frac{\delta\mu_n}{\mu}\right) \right] \,,\label{eq:a:v_neutrons}\\
\Vect{v_{ad}} = \frac{\xi}{\gamma_{cn}n_c n_n}\left[b^2\left(\Curl\Vect{B}\right)\times\Vect B-n_c\mu\Grad \left(\frac{\delta\mu_c}{\mu}\right) \right] \label{eq:a:v_ambipolar}\,,\\
\delta\mu_c = K_{cc}\delta n_c \label{eq:a:dmuc}\,,\\
\delta\mu_n = K_{nn}\delta n_n \label{eq:a:dmun}\,,
\end{gather}
where distances have been normalized to the core radius $R$, number densities $n_n$ and $n_c$ are in units of $n_{c0}$, $K_{nn}$ and $K_{cc}$ are in units of $K_{cc0}=K_{cc}(r=0)$, $\gamma_{cn}$ is in units of $\gamma_{cn0}=\gamma_{cn}(r=0)$, time has been normalized to $t_{0}=\xi R^2 \gamma_{cn0}/K_{cc0}$, chemical potentials are in units of $K_{cc0}n_{c0}$, $\zeta$ is in units of $\gamma_{cn0}n_{c0}$, velocities are normalized to $R/t_{0}$, and the magnetic field is in units of $B_0$ (the root mean square of the initial magnetic field in the volume of the star).
$\alpha$ and $\beta$ are in units of $R^2B_0$ and $RB_0$, respectively.
We control the strength of the magnetic field adjusting the parameter
\begin{equation}
b^2 \equiv \frac{B_0^2}{4\pi K_{cc0}n_{c0}^2} \label{eq:b2def}\,,
\end{equation}
which is of the order of the (very small) ratio between the magnetic pressure and the degeneracy pressure of the charged particles. Thus, for our NS model, $B_0 = n_{c0}\sqrt{4\pi K_{cc0}}\:b = 2\times 10^{18}b\, \text{G}$.

Hereafter, we take the value of the different time-scales at the center of the NS as reference values, which, in dimensionless units read, from the shortest to the longest,
\begin{gather}
t_{\zeta p} \equiv \left.\frac{K_{cc}n_c^2}{K_{nn}n_n^2}\right|_{r=0}\zeta x^2=\frac{\zeta x^2}{16.45} \,,\label{eq:t_zeta_p_a}\\
t_{\zeta g} \equiv \zeta x^2 \,,\label{eq:t_zeta_g_a}\\
t_{\zeta B} \equiv \frac{\zeta x^2}{ b^2}\,,\label{eq:t_zeta_B_a}\\
t_{ad} \equiv \frac{x^2}{b^2}  \,,
\end{gather}
where $x \equiv L/R$, and we take $x=1/4$, since it roughly fits the structure of the initial condition of our simulations.
In order to properly resolve all four time-scales in our simulations without having to use a prohibitively small time-step, we use unrealistically large values for the magnetic field (fixed by $b$) and $\zeta$, so the four time-scales are in the correct order of sizes, but get much closer to each other than in reality.

In order to evolve the magnetic field in the NS core, we upgraded the code described in paper I to include the motion of neutrons and the radial density gradients, so it evolves the set of equations~\eqref{eq:a:alpha}--\eqref{eq:a:dmun}.
This is done by discretizing the values of the variables over a staggered polar grid composed of $N_r$ points inhomogeneously distributed in the radial direction inside the core and $N_\theta$ points equally spaced in the polar direction, while the external multipolar expansion of $\alpha(r,\theta)$ is truncated to the first $N_\text{Exp}$ terms.
The numerical computation is done conservatively for the evolution of the toroidal magnetic field and the density perturbations of the charged particles and neutrons, using a finite-volume scheme.
The time derivative of the poloidal potential is computed using finite differences, and the system is evolved to second-order accuracy in time. This scheme guarantees that the $\Div\Vect{B}=0$ condition, as well as the total number of particles, are conserved at all times to machine precision.
Details on the numerical code can be seen in paper I.

\section{Results}
\label{sec:results}

\subsection{Dependence on the artificial friction}
\label{sec:results:z_comp}

\begin{figure}
	\centering
	\includegraphics[width=\linewidth]{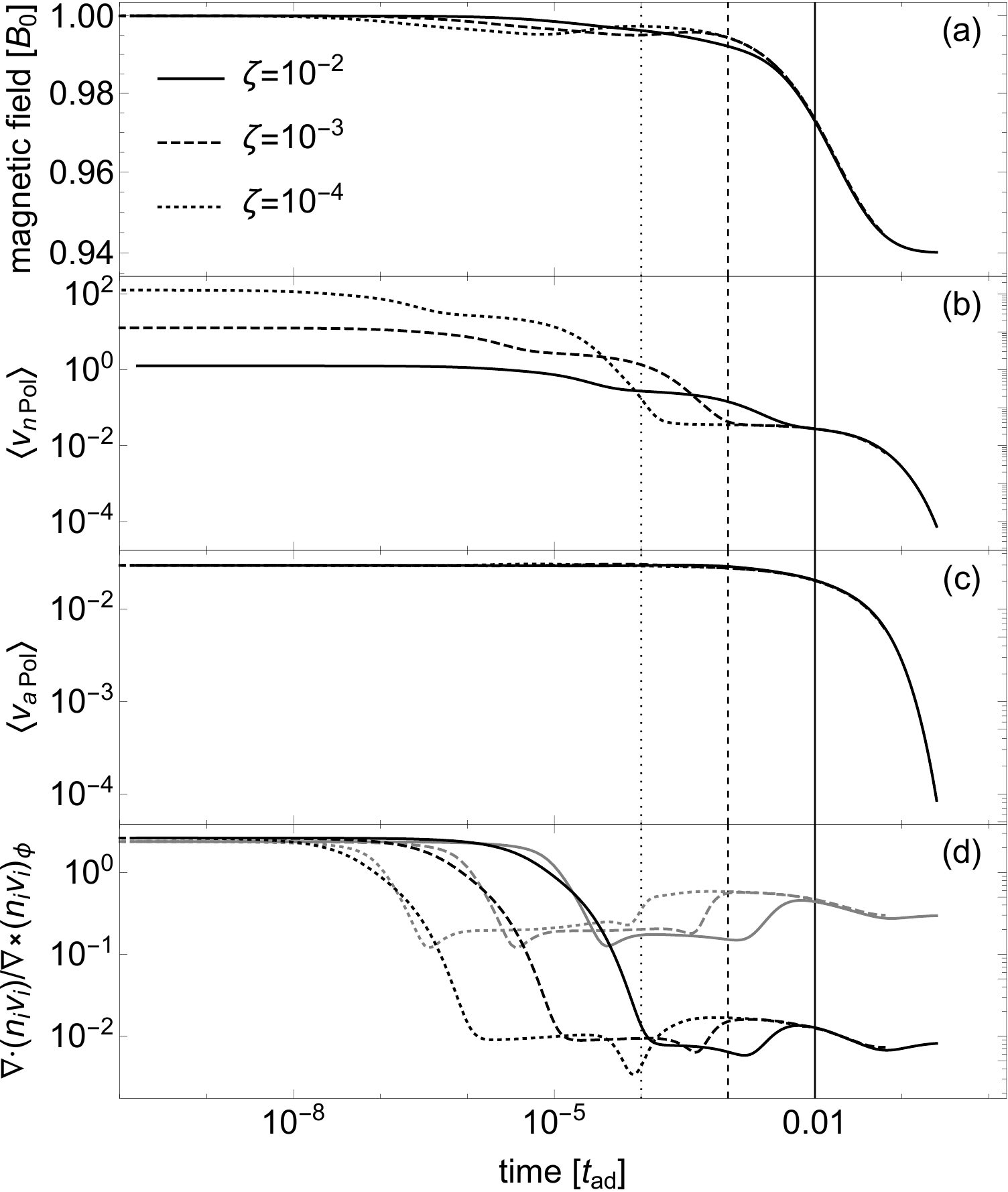}
	\caption{
		Comparison of the evolution of simulations using the initial condition from equation \eqref{eq:initial_alpha1} and three different values of $\zeta$ ($10^{-2}$, $10^{-3}$, and $10^{-4}$), which yield the following ratios between the different relevant time-scales at the center of the star: $t_{\zeta p}: t_{\zeta g} : t_{\zeta B} : t_{ad}=1 : 16.45 : 1645 : 1.645\times 10^5$
		(continuous line), $=1 : 16.45 : 1645 : 1.645\times 10^6$
		(dashed), and $=1 : 16.45 : 1645 : 1.645\times 10^7$
		(dotted), respectively.
		The vertical lines show the value of the time-scale $t_{\zeta B}$ for each simulation. We show the time evolution of (a) the magnetic field $\langle \Vect{B} \rangle$ in units of $B_0$, where $\langle .\rangle$ denotes rms in the volume of the core, (b) the neutron velocity $\langle \Vect{v_n} \rangle$, (c) the ambipolar diffusion velocity $\langle \Vect{v_{ad}} \rangle$, and (d) the ratio $\langle \Div (n_i\Vect{v_{i}} )\rangle/\langle [\Curl( n_i\Vect{v_{i}} )]_\phi\rangle$ for neutrons ($i=n$, black lines), and charged particles ($i=c$, gray lines).
	}
	\label{fig:z_comp}
\end{figure}

In our model, the fictitious friction force is a device to allow the particle densities and the magnetic field configuration to reach a hydro-magnetic equilibrium on time-scales $\lesssim t_{\zeta B}$ shorter than the ambipolar diffusion time $t_{ad}$, but close enough to the latter for the numerical simulation to be feasible.
Thus, the friction parameter $\zeta$ needs to be chosen to satisfy this compromise, and its value should not affect the long-term evolution of the magnetic field on scales $\sim t_{ad}$. In order to estimate its optimal value, we compare three simulations that differ only by their value of $\zeta$.
Since the total integration time is proportional to $t_{ad}/t_{\zeta p}=16.45/\zeta b^2$, decreasing the values of $\zeta$ and $b$ increases the integration time very quickly. Therefore, to keep it manageable, we perform this test with an unrealistically large value of $b^2$, namely, $10^{-2}$.
For $\zeta$, we take the values $\zeta_1=10^{-2}$, $\zeta_2=10^{-3}$, and $\zeta_3=10^{-4}$, which yield ratios $t_{ad}/t_{\zeta B}=10^{2}$, $10^{3}$, and $10^{4}$, respectively.

As initial condition, we choose the purely poloidal magnetic field generated by the potential 
\begin{equation}
\alpha_1(r,\theta)=\alpha_{01}r^2\left( 1 -\frac 65r^2 +\frac 37r^4 \right)\sin^2\theta\,,\label{eq:initial_alpha1}
\end{equation}
where $\alpha_{01}=1.336$ is a normalization constant, fixed by the condition $\langle B \rangle=1$ on the normalized magnetic field, where $\langle .\rangle$ denotes rms in the volume of the core.
We chose this configuration, as it is one of the simplest analytic expressions matching all the constraints required for $\Vect B$, and it has been widely used in the literature \citep{Akgun2013,Passamonti2017,Ofengeim2018}.
As in the discussion of \S~\ref{sec:model:timescales:1}, we take the initial conditions $\delta n_n=\delta n_c=0$ for the density perturbations. 

Our results are summarized in Fig.~\ref{fig:z_comp}. Panel (a) shows that during the early stages ($\propto t_{\zeta B}$), there is a small adjustment of the magnetic field.
This is expected, as growing the density perturbations up to $|\delta n_i|/n_i\sim b^2$ (for $i=n,c$) implies spatial displacements of a fraction $\sim b^2$ of the core radius thus the magnetic field strength is expected to be reduced by a similar fraction.
We also see that, while the early dynamics is dominated by the artificial friction, the three simulations converge to the same curve at $t\sim t_{\zeta B}\propto\zeta$.
This can be seen more clearly in panel (b).
The convergence of $\Vect{v_n}$ for simulations with very different values of $\zeta$ confirms that the scheme is indeed yielding the expected results, namely $f_\zeta\propto\zeta$, thus reproducing the correct ``physical'' neutron velocity on time-scales $t\sim t_{\zeta B}$, independent of the value used for $\zeta$ (as long as it is sufficiently small).
The good agreement of the three curves after this time suggests that all three values chosen for $\zeta$ are adequate for our purpose.
To be on the safe side, we will use $\zeta\sim 10^{-3}$ to perform long-term simulations (reaching $t\sim t_{ad}$), as it still yields manageable integration times.

It is usually assumed in the literature that during the long-term evolution of the field, the time derivatives in the continuity equations are negligible \citep{Goldreich1992a,Gusakov2017}.
Since in our approach we are explicitly evolving the fluid perturbations, it is interesting to see if our scheme confirms this assumption.
Neglecting time derivatives in equations \eqref{eq:continuity_n} and \eqref{eq:continuity_c} implies that both $n_n\Vect{v_n}$ and $n_c\Vect{v_c}$ will be perfectly solenoidal during the long-term evolution.
The simulation shown in the next subsection (see Figs.~\ref{fig:z1e-4_snapshot}[d] and [e] at $t=t_{\zeta B}$) suggests that $\Vect{v_n}$ and $\Vect{v_c}=\Vect{v_n}+\Vect{v_{ad}}$ are indeed mostly solenoidal.
This becomes even clearer from Fig.~\ref{fig:z_comp}(d). 
For neutrons,
\begin{equation}
    \frac{\Div(n_n\Vect{v_{n}})}{[\Curl(n_n\Vect{v_{n}})]_\phi}\sim\frac{\delta n_n/t_{\zeta B}}{n_n v_{n}/L}\sim \frac{\delta n_n}{n_n} \sim \frac{B^2}{4\pi K_{nn}n_n^2}\sim\frac{b^2}{16}\,,
\end{equation}
which roughly agrees with our results. Similarly, for charged particles, $\Div(n_c\Vect{v_{c}})/[\Curl(n_c\Vect{v_{c}})]_\phi \sim \delta n_c/n_c\sim b^2$.
These estimates are in agreement with the arguments used in the literature to neglect the time derivatives \citep{Goldreich1992a,Gusakov2017}.
As $\delta n_i/n_i$ scales with $b^2$ ($i=n,c$), it is expected that both approaches are equivalent when running simulations using realistic values of $b$ ($10^{-20}<b^2<10^{-4}$), however this is not possible at the moment, due to computing time constraints.

It remains to be seen if the fictitious force is indeed negligible during the long-term evolution of the field, and if our estimates for the different time-scales of evolution are in agreement with our results. This will be addressed in the following section.

\subsection{Hydro-magnetic evolution}
\label{sec:results:evolution}

\subsubsection{Reaching hydro-magnetic quasi-equilibrium with a purely poloidal configuration}
\label{sec:results:evolution:poloidal}

\begin{figure}
	\centering
	\includegraphics[width=\linewidth]{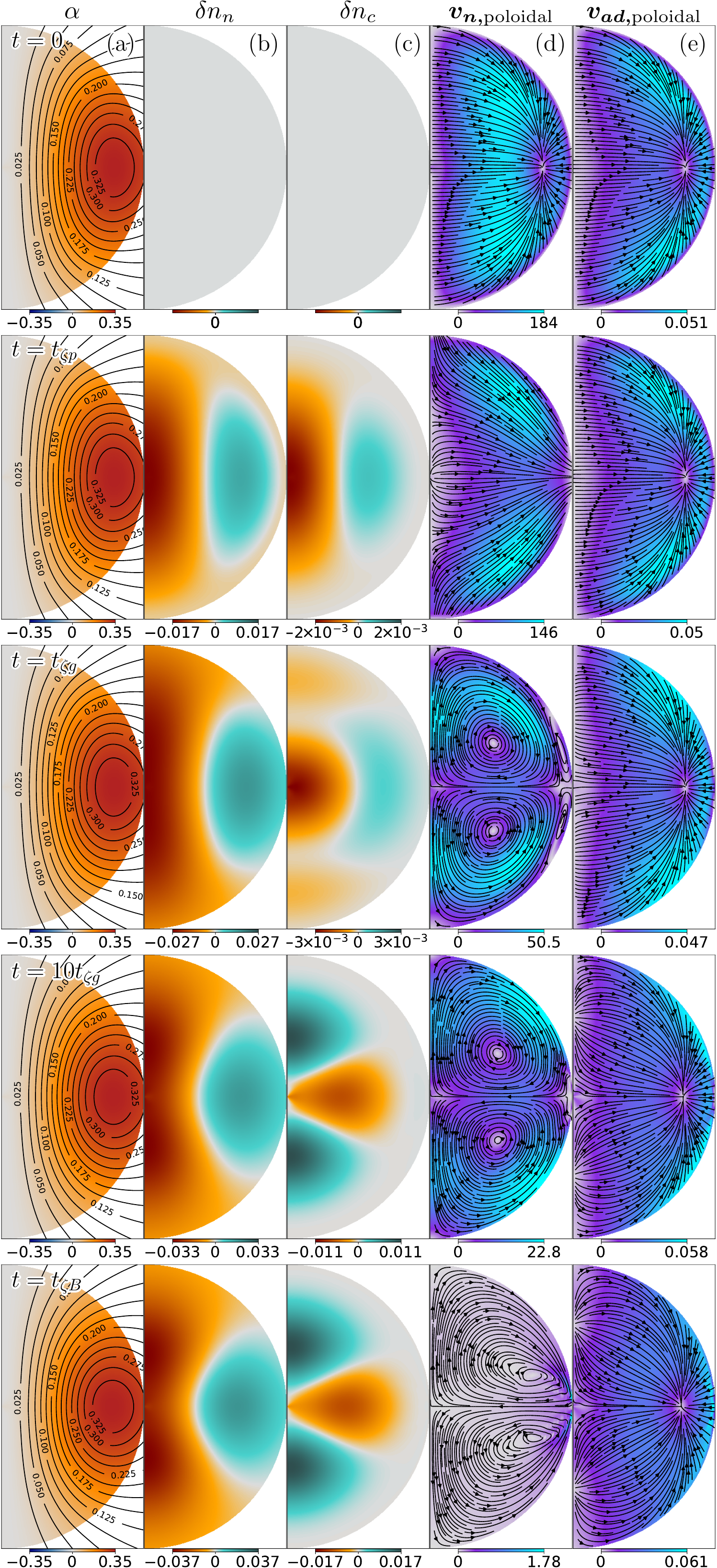}
	\caption{
		Evolution of the magnetic field discussed in \S~\ref{sec:results:evolution:poloidal}, with initial conditions set by $\alpha$ from equation \eqref{eq:initial_alpha1} and $\beta = 0$, and parameters $\zeta=10^{-4}$ and $b^2=10^{-2}$, yielding time-scales at the center of the star in proportions $t_{\zeta p} : t_{\zeta g}  : t_{\zeta B} : t_{ad}=1 : 16.45 : 1645 : 1.645\times 10^7$.
		We used a grid of $N_r=60$ radial steps and $N_\theta=91$ polar steps inside the core, as well as $N_{\text{Exp}}=27$ external multipoles. From left to right: (a) Configuration of the magnetic field, where lines represent the poloidal magnetic field and the colors the poloidal potential $\alpha$; (b) and (c) density perturbations $\delta n_n$ and $\delta n_c$, respectively, normalized to $n_{c0}$; (d) and (e) poloidal component of the neutron velocity, $\Vect{v_{n}}$, and ambipolar diffusion velocity, $\Vect{v_{ad}}$, where arrows represent the direction and colors the magnitude normalized to $R/t_0$. Rows correspond to different times: $t=0, t_{\zeta p}, t_{\zeta g}, 10t_{\zeta g}$, and $t_{\zeta B}$.
	}
	\label{fig:z1e-4_snapshot}
\end{figure}

In this section, we study the process in which the fictitious friction rearranges the particle densities and magnetic field to reach hydro-magnetic quasi-equilibrium, checking if the associated timescales agree with the analytical estimates of \S~\ref{sec:model:timescales}. For this purpose, we perform a simulation using the same initial condition as in the previous section (see equation~[\ref{eq:initial_alpha1}]).
The initial condition and some snapshots of its evolution are shown in Fig.~\ref{fig:z1e-4_snapshot}.

The short time-scales $t_{\zeta p}$ and $t_{\zeta g}$ (equations \ref{eq:t_zeta_p} and \ref{eq:t_zeta_g}, respectively) are associated to a quick adjustment of the densities in the presence of a magnetic force, nevertheless they do not explicitly depend on the magnetic field strength.
As they are much shorter than the time-scales for magnetic field evolution, the magnetic field remains essentially unchanged during this process.
Although in our simulations we use an unrealistically high value of $b$, the value chosen is still small enough to understand the dynamics of the fluid motion, since $t_{\zeta p}/t_{\zeta B}=6\times 10^{-4}$.
Also, here we are only interested in the dynamics on time-scales $\sim t_{\zeta B}$, thus we can use small values of $\zeta$ without changing the total integration time as the required number of time-steps will be $\propto t_{\zeta B}/t_{\zeta p}$, independent of $\zeta$. Therefore, we will focus, as suggested before, on the simulation with $\zeta=\zeta_3=10^{-4}$.

We can see that, as expected, at short times ($\sim t_{\zeta p}$), the magnetic field pushes the fluid, which grows density perturbations $\delta n_n(r,\theta)$ and $\delta n_c(r,\theta)$ with similar structure, in particular the same sign, as would happen in sound waves (p-modes, in asteroseismology jargon).
At later times ($\sim t_{\zeta g}$), although by construction the two species are still mostly moving together (as $\Vect{v_{n}} \gg \Vect{v_{ad}}$, since we are still out of quasi-equilibrium, implying $\Vect{v_n} \approx \Vect{v_c}$), their different background density profiles ($n_c(r)/n_n(r)\neq\mathrm{constant}$) together with a non-trivial velocity field $\Vect{v}_n(r,\theta)$ allow their continuity equations~(\ref{eq:a:continuity_n}) and (\ref{eq:a:continuity_c}) to create density perturbations not necessarily of the same signs (as they would have in so-called ``gravity waves'' or ``g-modes'', e.~g., \citealt{Reisenegger1992}), allowing their different fluid forces to jointly balance both components of the poloidal magnetic force (see panels at $t=t_{\zeta g}$ and $t=10t_{\zeta g}$).
This leads joint radial motions to be opposed by strong buoyancy forces \citep{Pethick1992,Goldreich1992a}.
Thus, the two components jointly behave as a single, stably stratified, non-barotropic fluid. 

\begin{figure}
	\centering
 \includegraphics[width=\linewidth]{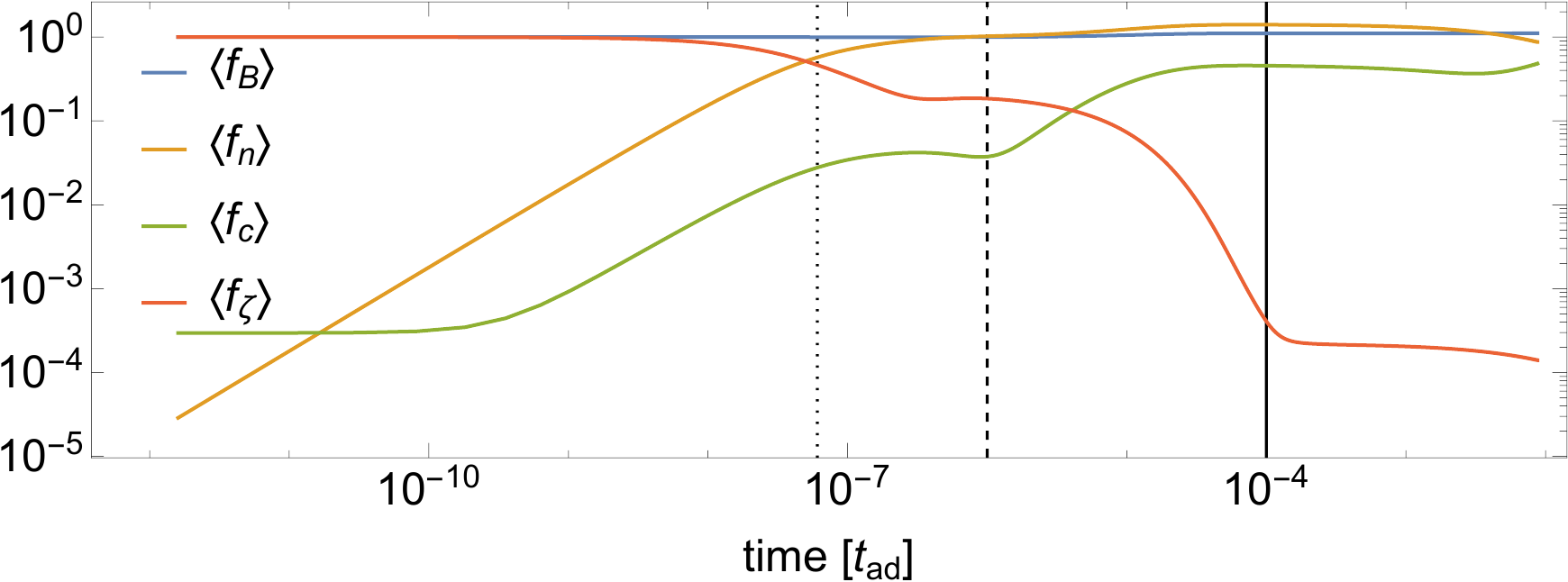}
	\caption{
	For the simulation of Fig.~\ref{fig:z1e-4_snapshot}: Time evolution of the force densities $\langle \Vect{f_B} \rangle$, $\langle \Vect{f_n} \rangle$, $\langle \Vect{f_c} \rangle$ and $\langle \Vect{f_\zeta} \rangle$,  all of them normalized by $\langle \Vect{f_B}(t=0)\rangle$, where $\langle .\rangle$ denotes rms in the volume of the core. 
	Time is in units of $t_{ad}$. The vertical lines show, from left to right, the values of the time-scales $t_{\zeta p}$, $t_{\zeta g}$, and $t_{\zeta B}$.
	}
	\label{fig:z1e-4_forces}
\end{figure}

\begin{figure}
	\centering
	\includegraphics[width=\linewidth]{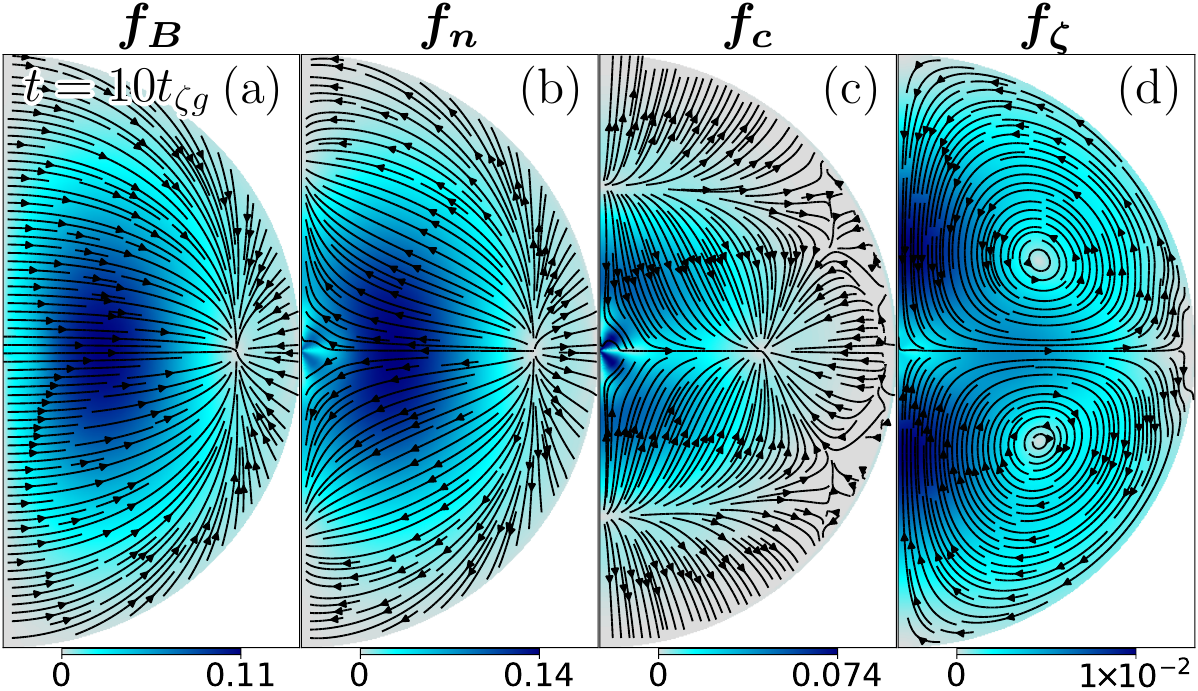}
	\caption{
    From left to right: Snapshot of (a) $\Vect{f_B}$, (b) $\Vect{f_n}$, (c) $\Vect{f_c}$, and (d) $\Vect{f_\zeta}$ for the simulation of Fig.~\ref{fig:z1e-4_snapshot} at $t=10t_{\zeta g}$. Arrows represent the direction of the forces at each point, and colors their magnitudes.
	}
	\label{fig:z1e-4_snapshot_forces}
\end{figure}

The strength of the magnetic and fluid forces throughout the simulation can be seen in Fig.~\ref{fig:z1e-4_forces}.
At early stages ($\sim t_{\zeta p}$), the magnetic force is partially balanced by the fluid force of the neutrons, while the contribution of the charged particles is much smaller.
This is because at this stage the charged particles and neutrons at any point in the star are jointly compressed or expanded by the magnetic force (with $\delta n_c(r,\theta)/n_c(r)\sim\delta n_n(r,\theta)/n_n(r)$).
Since there are many more neutrons than charged particles present, they represent the main contribution to the pressure of the fluid.
At later times ($\sim t_{\zeta g}$), the neutron and charged-particle density perturbations become very different and the latter also contribute substantially to the fluid force.
At first, it may be surprising that the two fluid forces do not act in the same direction, opposite to the magnetic force.
Instead, they grow very different density perturbations that, acting together, balance the magnetic force (see the red line).
It is clear from \S~\ref{sec:model:timescales:1} that to reach this kind of force balance, density perturbations of similar magnitude (but not necessarily of the same sign) are required.
The relative smallness of $f_\zeta$ in Fig.~\ref{fig:z1e-4_snapshot_forces}(d), shows that this force balance is being reached.
Nevertheless, as the magnetic flux is slowly being carried by the motion of the fluid, a perfect force balance cannot be reached, as seen in Fig.~\ref{fig:z1e-4_forces}: After their fast initial growth (in a time-scale $\sim t_{\zeta g}$), the fluid forces reach a plateau.
However, as the magnetic field evolves (on a much slower time-scale), the particles have to keep up, thus evolving at the same rate as the magnetic field, until the latter reaches a new quasi-equilibrium configuration at times $\sim t_{\zeta B}$.
At this point, all forces are close to balancing, and the fictitious friction becomes negligible compared to all of the physical forces ($|f_\zeta|/\text{min}\{|f_B|,|f_c|,|f_n|\}\propto\zeta/\gamma_{cn}n_c$). Thus, at this point the fictitious force plays no significant role in the evolution of the field, other than giving us a way of computing the neutron velocity field needed to maintain the quasi-equilibrium condition $f_\zeta\sim 0$.
It is also worth noting that the topology of the velocity fields at this point is consistent with the results from \citet{Ofengeim2018} (see Fig.~\ref{fig:z1e-4_snapshot}[d] and [e] at $t=t_{\zeta B}$), although a quantitative comparison is not possible since our toy-model equation of state does not consider entrainment.

\subsubsection{Long-term magnetic field evolution of a poloidal and toroidal configuration}
\label{sec:results:evolution:poloidal_toroidal}

In the previous sections, we studied simulations of a purely poloidal magnetic field, which is known to be unstable to non-axisymmetric perturbations \citep{Markey1973,Flowers1977,Marchant2011}.
Thus, in order to study the long-term evolution, we now consider a simulation for the much more realistic case of a field with poloidal and toroidal components.
This allows us to check whether the time-scales for the magnetic field evolution from section \ref{sec:model:timescales} (equations \ref{eq:t_zeta_B}, and \ref{eq:t_ad}) are in agreement with our numerical scheme, but mainly, it allows us to study the impact of the neutron motions on the long-term evolution of the magnetic field.

\begin{figure}
	\centering
	\includegraphics[width=\linewidth]{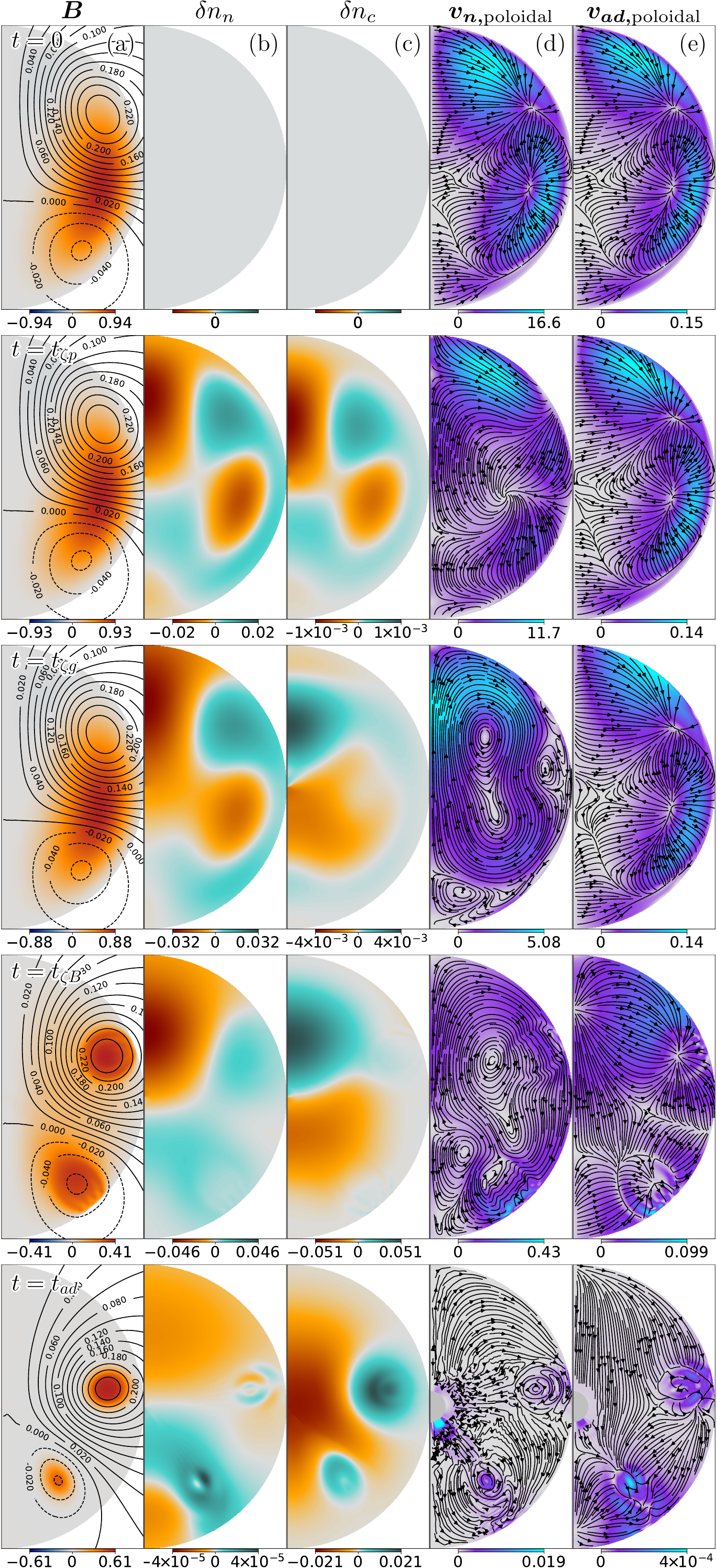}
	\caption{
	Evolution of the magnetic field discussed in \S~\ref{sec:results:evolution:poloidal_toroidal} with initial conditions given by equations \eqref{eq:initial_poltor1} and \eqref{eq:initial_poltor2} (so the poloidal component has 60\% of the initial internal magnetic energy), and parameters $\zeta=3\times10^{-3}$ and $b^2=10^{-2}$. Thus, the time-scales at the center of the star satisfy $t_{\zeta p} : t_{\zeta g}  : t_{\zeta B} : t_{ad}=1 : 16.45 : 1645 : 548500 $. We used a grid of $N_r=60$ radial steps and $N_\theta=91$ polar steps inside the core, as well as $N_{\text{Exp}}=27$ external multipoles. From left to right: (a) Configuration of the magnetic field, where lines represent the poloidal magnetic field (labeled by the magnitude of $\alpha$) and colors the toroidal potential $\beta$; (b) and (c) density perturbations $\delta n_n$ and $\delta n_c$, respectively, both notmalized to $n_{c0}$; (d) and (e) poloidal component of the neutron velocity, $\Vect{v_n}$, and ambipolar diffusion velocity, $\Vect{v_{ad}}$, where arrows represent the direction and colors the magnitude normalized to $R/t_0$. Rows correspond to different times: $t=0, t_{\zeta p}, t_{\zeta g}, t_{\zeta B}$, and $t_{ad}$.
	}
	\label{fig:poloidal_toroidal_snapshot}
\end{figure}

In addition to the dipolar potential $\alpha_1$ defined in \S~\ref{sec:results:z_comp}, we now consider an additional quadrupolar poloidal potential, 
\begin{equation}
\alpha_2(r,\theta)=\alpha_{02}r^3\left( 1 -\frac {10}7r^2 +\frac 59r^4 \right)\sin^2\theta\cos\theta\,,\label{eq:initial_alpha2}
\end{equation}
where $\alpha_{02}=1.239$ is a normalization constant, fixed by the condition $\langle B_{\text{Pol}} \rangle=1$, and a toroidal potential 
\begin{equation}
\beta_1(r,\theta)=\beta_{01}r^5(1 -r)^2\sin^2\theta \sin(\theta-\pi/5)\,,\label{eq:initial_beta1}
\end{equation}
where $\beta_{01}=112.546$ is fixed by the condition $\langle B_{\text{Tor}} \rangle=1$. We choose the initial condition as
\begin{gather}
\alpha(r,\theta) = \sqrt{0.18}\alpha_1(r,\theta)+\sqrt{0.42}\alpha_2(r,\theta)\,,\label{eq:initial_poltor1}\\
\beta(r,\theta) = \sqrt{0.4}\beta_1(r,\theta) \label{eq:initial_poltor2} \,,
\end{gather}
so the toroidal magnetic field has 40\% of the initial internal magnetic energy.
The artificial friction is set to $\zeta=3\times10^{-3}$, which yields an initial ratio $\langle v_n\rangle/\langle v_{ad}\rangle\sim 130$.

\begin{figure}
	\centering
	\includegraphics[width=\linewidth]{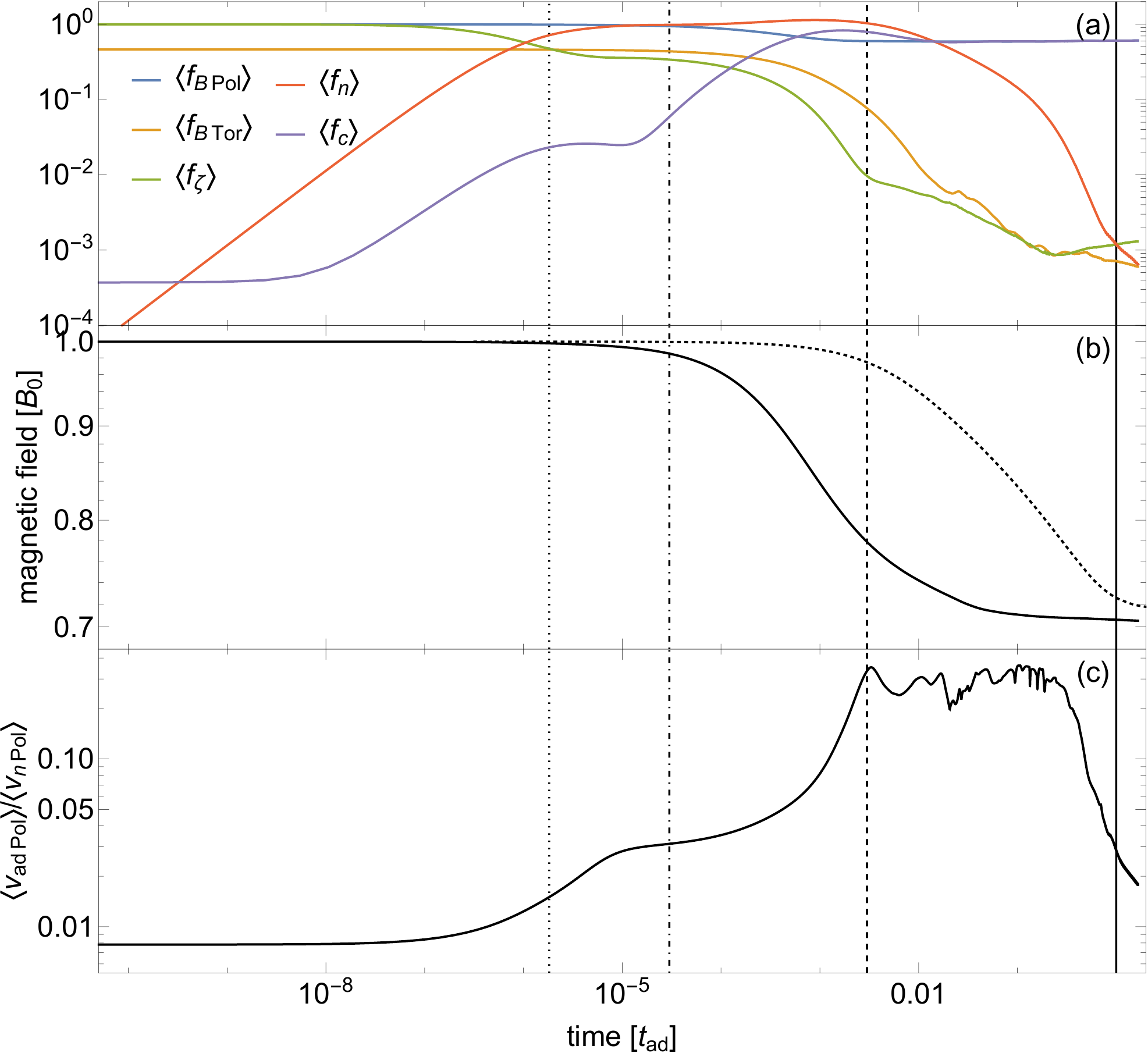}
	\caption{For the simulation of Fig.~\ref{fig:poloidal_toroidal_snapshot}: Time evolution of (a) the rms force densities $\langle \Vect{f_{B\,\text{Pol}}} \rangle$, $\langle \Vect{f_{B\,\text{Tor}}} \rangle$, $\langle \Vect{f_n} \rangle$, $\langle \Vect{f_c} \rangle$, and $\langle \Vect{f_\zeta} \rangle$, all normalized by $\langle \Vect{f_{B\,\text{Pol}}}(t=0)\rangle$, where $\langle .\rangle$ denotes rms in the volume of the core;
	(b) the rms magnetic field $\langle \Vect{B} \rangle$ (in units of its initial value $B_{0}$) in the NS core. The dotted line corresponds to $\langle \Vect{B} \rangle$ on a simulation using the same initial condition, but taking $\zeta\rightarrow\infty$ (fixed neutrons); and (c) $\langle v_{ad\,\text{Pol}}\rangle/\langle v_{n\,\text{Pol}}\rangle$.
	For all panels, time is in units of $t_{ad}$. The vertical lines show, from left to right, the time-scales $t_{\zeta p}$, $t_{\zeta g}$, $t_{\zeta B}$, and $t_{ad}$.
	}
	\label{fig:poloidal_toroidal_prom}
\end{figure}

The initial configuration, as well as some snapshots of its evolution, are shown in  Fig.~\ref{fig:poloidal_toroidal_snapshot}.
We see how at $t=t_{\zeta p}$ the density perturbations grow together following $\delta n_c/n_c\sim\delta n_n/n_n$, while at $t=t_{\zeta g}$ they are evolving towards a state in which their signs are not correlated, thus reproducing the results from the previous section.
This can also be observed in Fig.~\ref{fig:poloidal_toroidal_prom}(a), which shows the rms of the different fluid forces throughout the simulation.
During the first stages of the evolution ($t\sim t_{\zeta p}$--$t_{\zeta g}$), the magnetic force is partially balanced by the fluid forces, which are dominated by the neutrons.

In the fourth row of Fig.~\ref{fig:poloidal_toroidal_snapshot} (at $t=t_{\zeta B}$), it can be seen how the magnetic field unwinds, eliminating most of the toroidal component, except in the regions of closed poloidal field lines. 
As discussed in paper I, such ``twisted-torus'' configurations are expected in axially symmetric hydro-magnetic equilibria, because there is no toroidal pressure gradient or gravitational force available to balance a toroidal component of the Lorentz force.
This implies the restriction $\beta=\beta(\alpha)$ on the potentials, which is also verified in the figure (see also Fig.~\ref{fig:poloidal_toroidal_snapshot2}).
This state corresponds to a non-barotropic quasi-equilibrium, as both charged particles and neutrons, which have different fluid forces, are contributing to balance the magnetic force.
This can be seen more clearly in Fig.~\ref{fig:poloidal_toroidal_prom}(a), where the poloidal magnetic force and fluid forces are all of the same order, but the poloidal force imbalance $\langle \Vect{f_{\zeta,\text{Pol}}} \rangle$ is much smaller, meaning that in the poloidal component there is a balance between the magnetic force and the induced fluid forces.
In the toroidal component, $\Vect{f_{\zeta,\text{Tor}}}=\Vect{f_{B,\text{Tor}}}$ always, but at $t_{\zeta B}$ this force has become much smaller than its initial value. Thus, the configuration at $t\sim t_{\zeta B}$ is very close to a (non-barotropic) hydro-magnetic quasi-equilibrium where even the toroidal component of friction force has become small.

On a much longer time-scale, this quasi-equilibrium is slowly eroded by ambipolar diffusion.
Charged particles, pushed by the magnetic force, carry the magnetic flux relative to the neutrons until the magnetic force can be balanced by the force due to the density perturbations of the charged particles alone, choking the motion.
This can be seen in the last row of Fig.~\ref{fig:poloidal_toroidal_snapshot} ($t=t_{ad}$), where the ambipolar velocity is much smaller than at $t_{\zeta B}$.
Also, $|\delta n_c|/n_c\gg |\delta n_n|/n_n$, confirming that the magnetic force is mostly balanced by the charged particles.
This is further supported by Fig.~\ref{fig:poloidal_toroidal_prom}(a), which shows a transition from $t=t_{\zeta B}$, where the magnetic force is balanced by the combined force from charged particles and neutrons, to $t=t_{ad}$, where it is balanced mostly by the charged particles, while the contribution from the neutrons becomes negligible.

However, this does not mean that the neutron motion is irrelevant to the long-term evolution.
While ambipolar diffusion pushes the charged particles relative to the neutrons, all components (neutrons, charged particles, and the magnetic field) have to keep adjusting to successive quasi-equilibria, thus inducing a slow macroscopic motion of the fluid.
This motion (controlled by the drag force between neutrons and charged particles) gives the charged particles an extra push, shortening the time-scale for the long-term magnetic evolution ($\sim L/v_c$).
This can be seen in Fig.~\ref{fig:poloidal_toroidal_prom}(b), where the decay of the magnetic field in the simulation with mobile neutrons is much faster than in a simulation with the same initial condition, but with fixed neutrons.

As discussed in \S~\ref{sec:model:timescales}, we expect that in the quasi-stationary state $|\partial \delta n_c/\partial t|$ is much smaller than both $|\Div\left(n_c\Vect{v_{ad}}\right)|$ and $|\Div\left(n_c\Vect{v_n}\right)|$.
Thus, the irrotational components of $n_c\Vect{v_n}$, and $n_c\Vect{v_{ad}}$ are of the same order. 
However, as pointed out in \S~\ref{sec:results:z_comp}, both vector fields are dominated by their solenoidal component, and Fig.~\ref{fig:poloidal_toroidal_prom}(c) shows that, during the long-term evolution from $t_{\zeta B}$ to $t_{ad}$, $v_n$ becomes larger than $v_{ad}$ by a factor $\sim 4$. Similar fractions were found for different simulations in our tests. This is in agreement with the results of \citet{Ofengeim2018}. In Fig.~\ref{fig:poloidal_toroidal_prom}(c) we also see that, towards the end on the simulation, the core is reaching equilibrium, thus $\Vect{v_{ad}}$ gets progressively smaller, faster than $\Vect{v_n}$.
Sadly, at this point the numerical noise starts to become dominant towards the center of the star, where the coordinate system is singular (see Fig.~\ref{fig:poloidal_toroidal_snapshot}[d] at $t=t_{ad}$).

\subsection{Grad-Shafranov equilibria}
\label{sec:results:GS}

\begin{figure}
	\centering
	\includegraphics[width=\linewidth]{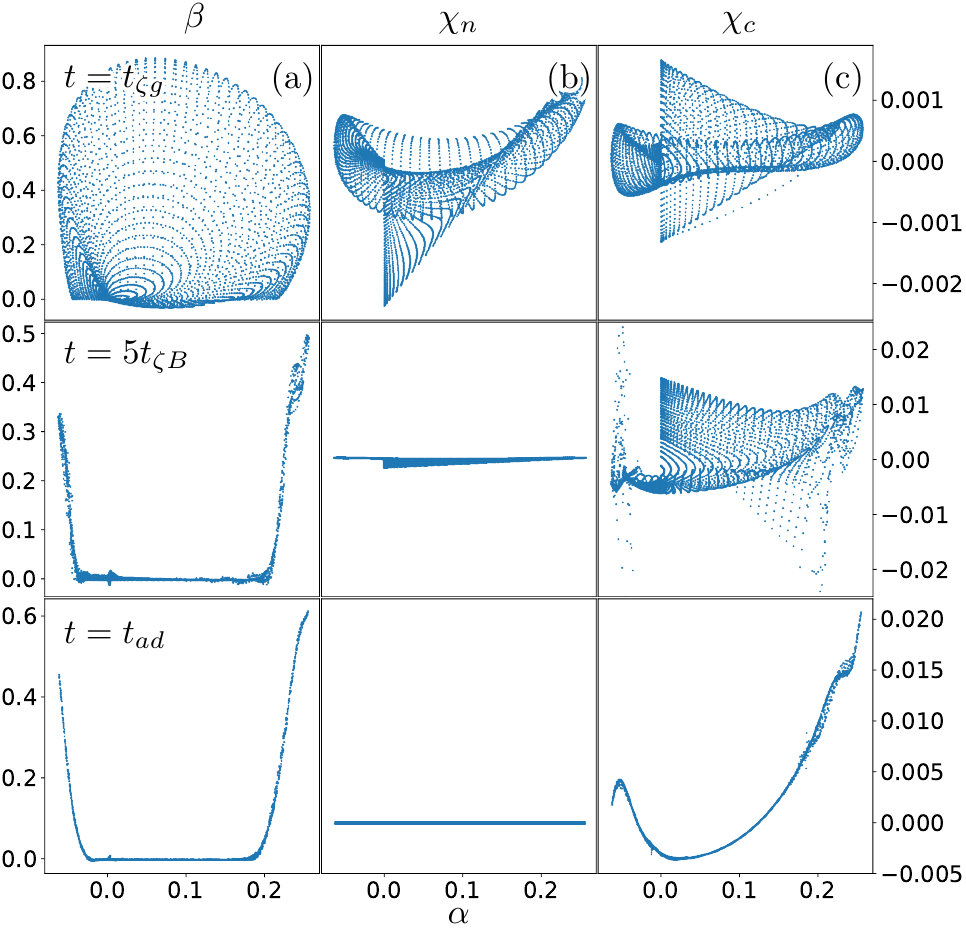}
	\caption{For the simulation of Fig.~\ref{fig:poloidal_toroidal_snapshot}: 
	Scatter plot of $\alpha$ versus (a) $\beta$, (b) $\chi_n$, and (c) $\chi_c$, showing all of the grid points at $t=t_{\zeta g}$, $t=5t_{\zeta B}$, and $t=t_{ad}$, respectively. The axis label on the left side of the figure corresponds to plot (a), while that on the right corresponds to (b) and (c).
	}
	\label{fig:poloidal_toroidal_snapshot2}
\end{figure}

We saw that, within a few $t_{\zeta B}$, the NS core reaches a hydro-magnetic quasi-equilibrium in which all forces are close to balance. As described in the previous section, this implies that the toroidal magnetic force must vanish, since there are no other azimuthal forces available to balance it, so
\begin{equation} 
\Grad\alpha\times\Grad\beta=0, \label{eq:Galpha_times_Gbeta}
\end{equation}
requiring (at least locally) a relation $\beta=\beta(\alpha)$.
This can be verified in Fig.~\ref{fig:poloidal_toroidal_snapshot2}, where at early times ($\sim t_{\zeta g}$) there is no clear relation between the variables, while at
$\sim t_{\zeta B}$ there is an evident dependence of $\beta$ on $\alpha$. However, at that time, no similar relations are found for other variables, such as $\delta n_i$ or $\delta\mu_i$ ($i=n,c$).

At later times ($\sim t_{ad}$), the velocities become much smaller than their initial value, suggesting that the system is approaching an equilibrium state in which both $\Vect{v_n}=0$ and $\Vect{v_{ad}}=0$.
The first condition implies $\Vect{f_{B}}+\Vect{f_{n}}+\Vect{f_{c}}=0$, while the second implies $\Vect{f_{B}}+\Vect{f_{c}}=0$ (see equations [\ref{eq:v_neutrons}] and [\ref{eq:v_ambipolar}]). As the toroidal magnetic force is already negligible at this point, this requires 
\begin{gather}
\frac{\Vect{J}}{c}\times\Vect{B}-n_c\mu\Grad \left(\frac{\delta\mu_c}{\mu}\right)=0\,,\label{eq:charged equilibrium}\\
\Grad \left(\frac{\delta\mu_n}{\mu}\right)=0 \,, \label{eq:neutron equilibrium}
\end{gather}
implying that the neutrons are in diffusive equilibrium (unperturbed by the presence of the magnetic field), while the forces due to the charged-particle density perturbations must balance the Lorentz force.
This, together with equation~\eqref{eq:Galpha_times_Gbeta}, implies that $\chi_c\equiv\delta\mu_c/\mu$ must also be a function of $\alpha$ (while $\chi_n\equiv\delta\mu_n/\mu$ must be uniform).
This can be verified in Fig.~\ref{fig:poloidal_toroidal_snapshot2}, where at early times ($\sim t_{\zeta g}$ and $\sim t_{\zeta B}$) there is no clear relation between the variables, while at later times ($\sim t_{ad}$) there is an evident dependence on $\alpha$ for $\beta$ and $\chi_c$, while $\chi_n\approx 0$.

As discussed previously \citep{Reisenegger2009,Lander2009,Lander2012,Armaza2015,Castillo2017}, equation~(\ref{eq:Galpha_times_Gbeta}) can be used to rewrite equation~(\ref{eq:charged equilibrium}) as a ``Grad-Shafranov (GS) equation'' \citep{Grad1958,Shafranov1966}: 
\begin{equation}
\Sh\alpha+\beta\beta' + 4\pi r^2\sin^2\theta\, n_c(r)\mu(r)\chi_c' = 0 \,,\label{eq:GS}
\end{equation}
where 
\begin{equation}
\begin{split}
	\Sh & \equiv
	r^2\sin^2\theta\Div\left(\frac{\Grad }{r^2\sin^2\theta}\right) \label{eq:shafranov_op}\\
	&= \frac{\partial^2}{\partial r^2} + \frac{\sin\theta}{r^2} \frac{\partial}{\partial\theta}\left( \frac{1}{\sin\theta}\frac{\partial}{\partial\theta}\right) \,,
\end{split}
\end{equation}
is the ``GS operator'', $\beta$ and $\chi_c$ are functions of $\alpha(r,\theta)$, and primes denote derivatives with respect to $\alpha$. We emphasize that (as seen in Fig.~\ref{fig:poloidal_toroidal_snapshot2}) this equation is generally \emph{not} satisfied in the previous stage (at $t\sim t_{\zeta B}$), which is also a hydro-magnetic quasi-equilibrium, but with neutrons and charged particles collisionally coupled and thus jointly balancing the Lorentz force.

\begin{figure}
	\centering
	\includegraphics[width=\linewidth]{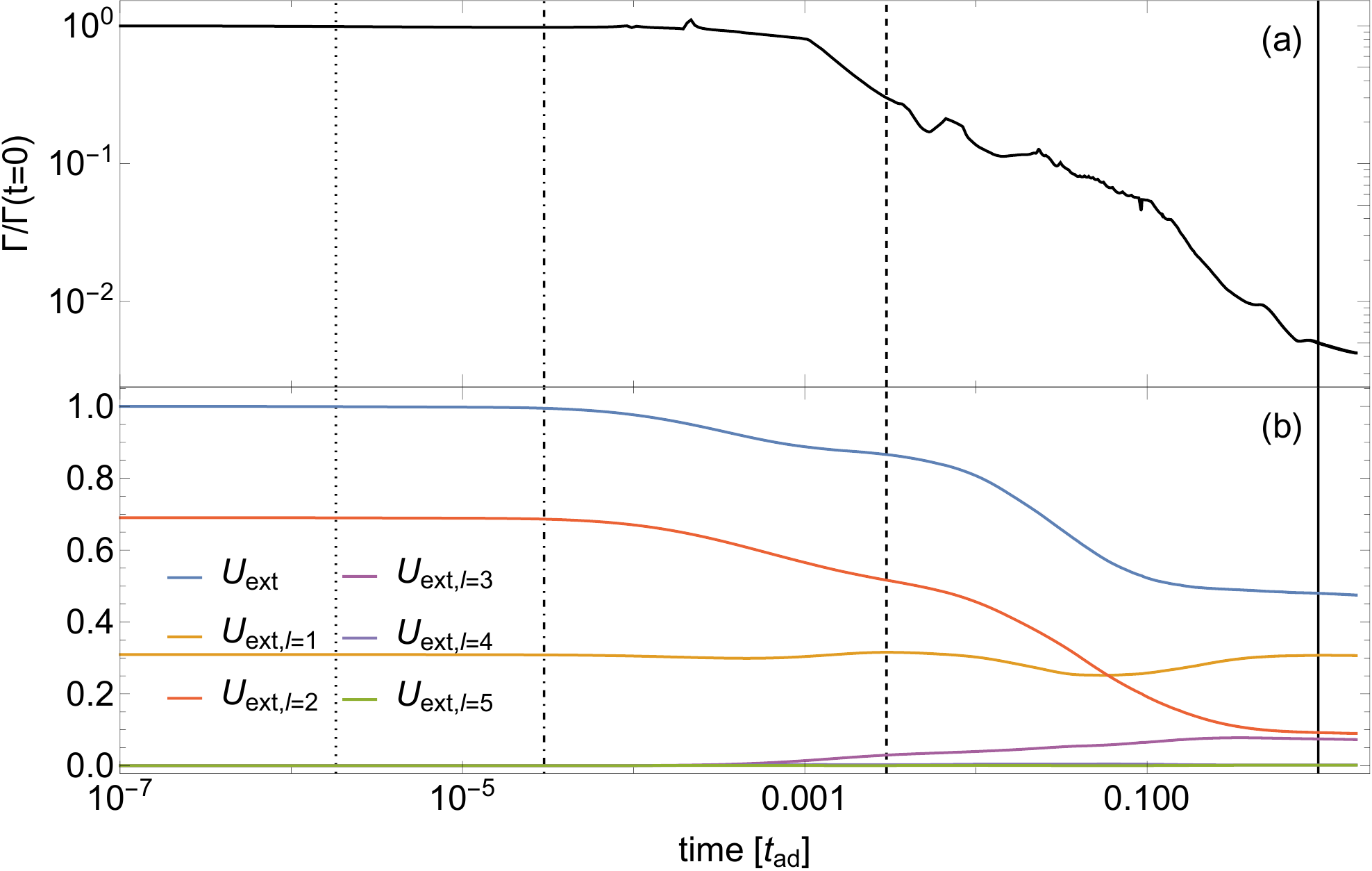}
	\caption{For the simulation of Fig.~\ref{fig:poloidal_toroidal_snapshot}: Time evolution of (a) the ``Grad-Shafranov integral'' $\Gamma$ (defined in equation~[\ref{eq:GS_prom}]) and (b) the magnetic energy stored in the $\ell$-th component of the external magnetic field (see equation~[\ref{eq:uext}]).
	The vertical lines show, from left to right, the time-scales $t_{\zeta p}$, $t_{\zeta g}$, $t_{\zeta B}$, and $t_{ad}$.
}
    \label{fig:sim_B_evo}
\end{figure}

In order to check if our simulations lead to configurations in which the GS equation~\eqref{eq:GS} is satisfied in the expected time-scale (a few $t_{ad}$), we evaluate the ``GS integral''
\begin{equation}
\Gamma=\frac{\int_V d^3x\left|\Sh\alpha+\beta\beta' + 4\pi r^2\sin^2\theta\, n_c\mu\chi' \right|}{V\max\left|\Sh\alpha\right|}\,, \label{eq:GS_prom}
\end{equation}
where the derivatives on $\alpha$ are computed from our simulations taking $\beta'=(\Grad\beta\cdot\Grad\alpha)/|\Grad\alpha|^2$, and $\chi_c'=(\Grad\chi_c\cdot\Grad\alpha)/|\Grad\alpha|^2$.
Fig.~\ref{fig:poloidal_toroidal_prom}(b) and Fig.~\ref{fig:sim_B_evo}(a) show that, in the time interval $t_{\zeta B}<t<t_{ad}$ the rms magnetic field strength (and thus square root of the internal magnetic energy) decays only from $78\%$ to $71\%$, the external magnetic energy decays from $87\%$ to $48\%$ of its initial value, and $\Gamma$ decays from $30\%$ to $0.5\%$ of its initial value, clearly showing that the magnetic field configuration approaches a GS equilibrium in $t\sim t_{ad}$.
Fig.~\ref{fig:sim_B_evo}(b) shows that during this process the poloidal magnetic field, which is initially a combination of dipolar ($\ell=1$) and quadrupolar ($\ell=2$) components, evolves towards a mostly dipolar configuration, where the quadrupolar component becomes progressively smaller, as an octupolar ($\ell=3$) component increases, reaching a similar magnitude in the final equilibrium.
It can also be seen that the contribution of higher multipoles is negligible.
Further simulations are required to check if this behavior is generic.
The results provided here may be a valuable resource for improving crustal magnetic field evolution models, which usually use boundary conditions relying on unphysical assumptions, such as crust confinement. 

\section{Conclusions}
\label{sec:conclusions}

In paper I, we studied the long-term evolution of the magnetic field of a NS through ambipolar diffusion, modeling its core as a charged-particle fluid of protons and electrons, which carries the magnetic flux through a motionless uniform background of neutrons.
In the present paper, we addressed two of the main shortcomings of that work by allowing neutrons to move as well as including different density gradients for neutrons and charged particles, thus accounting for their stable stratification.

We know that the star can reach a hydro-magnetic quasi-equilibrium many orders of magnitude faster than the magnetic field evolution time-scales.
To maintain quasi-equilibrium realistically during the evolution of the field would require following the propagation of sound, gravity, and Alfv{\'e}n waves.
Resolving such processes would make it impossible to simulate the long-term evolution of the field.
We have addressed this issue by adding a fictitious friction force to the equation of motion of the neutrons \citep{Hoyos2008}.
This force, which causes a neutron velocity proportional to the net force imbalance at each point, replaces the very small inertial terms present in the equations of motions, and provides a mechanism to maintain the hydro-magnetic quasi-equilibrium.\\

Our results can be summarized as follows:
\begin{enumerate}
\itemsep1em

\item The addition of the fictitious friction in our simulations was found to be a useful method to quickly reach and then maintain the hydro-magnetic quasi-equilibrium of the star without having to resolve the propagation of sound, gravity, and Alfv\'en waves, leading to the same kind of non-barotropic ``twisted torus'' quasi-equilibrium previously found in MHD simulations \citep{Braithwaite2004,Braithwaite2006a} in a time-scale $t_{\zeta B}$ that can be adjusted to make the simulations both doable and physically interesting.

\item We found good agreement between different expected time-scales involved in this process ($t_{\zeta p}$, $t_{\zeta g}$, and $t_{\zeta B}$), and simulations.
The long-term evolution of the magnetic field is not significantly affected by the fictitious friction force if the ratio between the time-scale in which the hydro-magnetic quasi-equilibrium is restored by this artificial friction and the time-scale of ambipolar diffusion is small enough ($t_{\zeta B}/t_{ad}\sim\zeta/\gamma_{cn}n_c\lesssim 10^{-3}$ in our simulations).

\item Relaxing the unrealistic assumption of fixed neutrons was found to impact the long-term evolution of the magnetic field in the following way:
Since the density profiles of neutrons and charged particles are different, joint motion of the two species is strongly constrained by buoyancy forces.
Thus, there will always be relative motions between species (i.e. ambipolar diffusion).
This process, in which the magnetic force pushes the charged particles and magnetic flux relative to the neutrons, slowly erodes the hydro-magnetic quasi-equilibrium; therefore neutrons, charged particles, and the magnetic field have to continuously adjust to restore it.
This motion makes the charged particles (and thus the magnetic flux) move faster than in the case of fixed neutrons, thus shortening the time-scale for magnetic evolution.
In our simulations we find that the speed of neutrons and charged particles can be $\sim 1$--$10$ times larger than their relative velocity, yielding a time-scale
\begin{equation}
t_{ad} \sim (0.3 - 3)\times 10^{3} \, \left(\frac{10^{15}\text{G}}{B}\right)^{2}\left(\frac{T}{10^8\text{K}}\right)^{2}\left(\frac{L}{1\text{km}}\right)^{2}\,\text{yr.}
\end{equation}
for the long-term evolution of the field.

\item The process described above leads to a final equilibrium state in which the magnetic force is balanced by the pressure and gravitational forces of the charged particles, while the neutron density perturbations become negligible, thus recovering the barotropic ``Grad-Shafranov equilibria'' from paper I. 
\end{enumerate}

The following two main caveats need to be addressed by future work:
\begin{enumerate}
\itemsep1em

\item It is likely that the magnetic equilibria found may become unstable if we relax the axial symmetry restriction, as previous works indicate that there are no stable hydro-magnetic equilibria in barotropic stars \citep{Braithwaite2009,Akgun2013,Mitchell2015}.
In that case ambipolar diffusion might dissipate all the magnetic flux, so the low magnetic field of millisecond pulsars might be explained by magnetic flux dissipation prior to accretion \citep{Cruces2019}.
However, if the charged particles also include muons, the charged fluid will no longer be barotropic, which may help to stabilize the magnetic field.

\item Neutron star matter is expected to be superfluid and superconducting at the temperatures of interest, which might have a major impact on the time-scales described here, as the coupling between the different species should, in principle, be many orders of magnitude smaller.
The dynamics of the field evolution might also be quite different from what we presented here, as the macroscopic effect of the dynamics and interactions between quantized neutron vortices and magnetic flux tubes needs to be taken into account.

\item In this work, we assumed a potential solution for the crustal and external magnetic field, thus assuming that the crust adjusts instantaneously to the evolution in the core. 
However, if the crust is highly conductive, this may not be the case as currents may keep circulating in the crust, leading to non-trivial configurations (see for instance \citealt{Cumming2004,Pons2007,Gourgouliatos2016,Akgun2018}).
Ultimately, this might indicate that it is the crust which sets the time-scale for the evolution of the external field, leading to more restrictive equilibrium configurations in the core.
\end{enumerate}

\section*{Acknowledgements}
We are very grateful to M. Gusakov, J. Pons, and the ANSWERS group in Santiago de Chile for useful discussions. This work was supported by FONDECYT projects 3180700 (F.C.), 1201582 (A.R.), and 1190703 (J.A.V.), and the Center for Astrophysics and Associated Technologies (CATA; CONICYT project Basal AFB-170002). J.A.V. thanks for the support of CEDENNA under CONICYT grant FB0807.

\section*{Data availability}
The data underlying this article will be shared on reasonable request to the corresponding author.




\bibliographystyle{mnras}
\bibliography{bibexport} 

\begin{thebibliography}{}
\makeatletter
\relax
\def\mn@urlcharsother{\let\do\@makeother \do\$\do\&\do\#\do\^\do\_\do\%\do\~}
\def\mn@doi{\begingroup\mn@urlcharsother \@ifnextchar [ {\mn@doi@}
  {\mn@doi@[]}}
\def\mn@doi@[#1]#2{\def\@tempa{#1}\ifx\@tempa\@empty \href
  {http://dx.doi.org/#2} {doi:#2}\else \href {http://dx.doi.org/#2} {#1}\fi
  \endgroup}
\def\mn@eprint#1#2{\mn@eprint@#1:#2::\@nil}
\def\mn@eprint@arXiv#1{\href {http://arxiv.org/abs/#1} {{\tt arXiv:#1}}}
\def\mn@eprint@dblp#1{\href {http://dblp.uni-trier.de/rec/bibtex/#1.xml}
  {dblp:#1}}
\def\mn@eprint@#1:#2:#3:#4\@nil{\def\@tempa {#1}\def\@tempb {#2}\def\@tempc
  {#3}\ifx \@tempc \@empty \let \@tempc \@tempb \let \@tempb \@tempa \fi \ifx
  \@tempb \@empty \def\@tempb {arXiv}\fi \@ifundefined
  {mn@eprint@\@tempb}{\@tempb:\@tempc}{\expandafter \expandafter \csname
  mn@eprint@\@tempb\endcsname \expandafter{\@tempc}}}

\bibitem[\protect\citeauthoryear{Akg{\"{u}}n, Reisenegger, Mastrano  \&
  Marchant}{Akg{\"{u}}n et~al.}{2013}]{Akgun2013}
Akg{\"{u}}n T.,  Reisenegger A.,  Mastrano A.,   Marchant P.,  2013, \mn@doi
  [MNRAS] {10.1093/mnras/stt913}, 433, 2445

\bibitem[\protect\citeauthoryear{Akg{\"{u}}n, Cerd{\'{a}}-Dur{\'{a}}n, Miralles
   \& Pons}{Akg{\"{u}}n et~al.}{2018}]{Akgun2018}
Akg{\"{u}}n T.,  Cerd{\'{a}}-Dur{\'{a}}n P.,  Miralles J.~A.,   Pons J.~A.,
  2018, \mn@doi [MNRAS] {10.1093/mnras/sty2669}, 481, 5331

\bibitem[\protect\citeauthoryear{Akmal, Pandharipande  \& Ravenhall}{Akmal
  et~al.}{1998}]{Akmal1998}
Akmal A.,  Pandharipande V.~R.,   Ravenhall D.~G.,  1998, \mn@doi [PhysRevC]
  {10.1103/PhysRevC.58.1804}, 58, 1804

\bibitem[\protect\citeauthoryear{Armaza, Reisenegger  \& Valdivia}{Armaza
  et~al.}{2015}]{Armaza2015}
Armaza C.,  Reisenegger A.,   Valdivia J.~A.,  2015, \mn@doi [ApJ]
  {10.1088/0004-637X/802/2/121}, 802, 121

\bibitem[\protect\citeauthoryear{Backer, Kulkarni, Heiles, Davis  \&
  Goss}{Backer et~al.}{1982}]{Backer1982}
Backer D.~C.,  Kulkarni S.~R.,  Heiles C.,  Davis M.~M.,   Goss W.~M.,  1982,
  \mn@doi [Nature] {10.1038/300615a0}, 300, 615

\bibitem[\protect\citeauthoryear{Bhattacharya}{Bhattacharya}{1991}]{Bhattacharya1991}
Bhattacharya D.,  1991, \mn@doi [Physics Reports]
  {10.1016/0370-1573(91)90064-S}, 203, 1

\bibitem[\protect\citeauthoryear{Bhattacharya}{Bhattacharya}{1995}]{Bhattacharya1995}
Bhattacharya D.,  1995, in Lewin W. H.~G.,  van Paradijs J.,   {E. P. J. van
  den Heuvel} eds, X-Ray Binaries. Cambridge University Press, p.~233

\bibitem[\protect\citeauthoryear{Braithwaite}{Braithwaite}{2009}]{Braithwaite2009}
Braithwaite J.,  2009, \mn@doi [MNRAS] {10.1111/j.1365-2966.2008.14034.x}, 397,
  763

\bibitem[\protect\citeauthoryear{Braithwaite \& Nordlund}{Braithwaite \&
  Nordlund}{2006}]{Braithwaite2006a}
Braithwaite J.,  Nordlund {\AA}.,  2006, \mn@doi [A{\&}A]
  {10.1051/0004-6361:20041980}, 450, 1077

\bibitem[\protect\citeauthoryear{Braithwaite \& Spruit}{Braithwaite \&
  Spruit}{2004}]{Braithwaite2004}
Braithwaite J.,  Spruit H.~C.,  2004, \mn@doi [Nature] {10.1038/nature02934},
  431, 819

\bibitem[\protect\citeauthoryear{Bransgrove, Levin  \& Beloborodov}{Bransgrove
  et~al.}{2018}]{Bransgrove2018}
Bransgrove A.,  Levin Y.,   Beloborodov A.,  2018, \mn@doi [MNRAS]
  {10.1093/mnras/stx2508}, 473, 2771

\bibitem[\protect\citeauthoryear{Castillo, Reisenegger  \& Valdivia}{Castillo
  et~al.}{2017}]{Castillo2017}
Castillo F.,  Reisenegger A.,   Valdivia J.~A.,  2017, \mn@doi [MNRAS]
  {10.1093/mnras/stx1604}, 471, 507

\bibitem[\protect\citeauthoryear{Cruces, Reisenegger  \& Tauris}{Cruces
  et~al.}{2019}]{Cruces2019}
Cruces M.,  Reisenegger A.,   Tauris T.~M.,  2019, \mn@doi [MNRAS]
  {10.1093/mnras/stz2701}, 490, 2013

\bibitem[\protect\citeauthoryear{Cumming, Arras  \& Zweibel}{Cumming
  et~al.}{2004}]{Cumming2004}
Cumming A.,  Arras P.,   Zweibel E.,  2004, \mn@doi [ApJ] {10.1086/421324},
  609, 999

\bibitem[\protect\citeauthoryear{Dommes \& Gusakov}{Dommes \&
  Gusakov}{2017}]{Dommes2017}
Dommes V.~A.,  Gusakov M.~E.,  2017, \mn@doi [MNRAS: Letters]
  {10.1093/mnrasl/slx011}, 467, L115

\bibitem[\protect\citeauthoryear{Elfritz, Pons, Rea, Glampedakis  \&
  Vigan{\`{o}}}{Elfritz et~al.}{2016}]{Elfritz2016}
Elfritz J.~G.,  Pons J.~A.,  Rea N.,  Glampedakis K.,   Vigan{\`{o}} D.,  2016,
  \mn@doi [MNRAS] {10.1093/mnras/stv2963}, 456, 4461

\bibitem[\protect\citeauthoryear{Flowers \& Ruderman}{Flowers \&
  Ruderman}{1977}]{Flowers1977}
Flowers E.,  Ruderman M.~A.,  1977, \mn@doi [ApJ] {10.1086/155359}, 215, 302

\bibitem[\protect\citeauthoryear{Gamow \& Schoenberg}{Gamow \&
  Schoenberg}{1941}]{Gamow1941}
Gamow G.,  Schoenberg M.,  1941, \mn@doi [PhysRev] {10.1103/PhysRev.59.539},
  59, 539

\bibitem[\protect\citeauthoryear{Glampedakis, Jones  \& Samuelsson}{Glampedakis
  et~al.}{2011}]{Glampedakis2011}
Glampedakis K.,  Jones D.~I.,   Samuelsson L.,  2011, \mn@doi [MNRAS]
  {10.1111/j.1365-2966.2011.18278.x}, 413, 2021

\bibitem[\protect\citeauthoryear{Goldreich \& Reisenegger}{Goldreich \&
  Reisenegger}{1992}]{Goldreich1992a}
Goldreich P.,  Reisenegger A.,  1992, \mn@doi [ApJ] {10.1086/171646}, 395, 250

\bibitem[\protect\citeauthoryear{Gourgouliatos \& Cumming}{Gourgouliatos \&
  Cumming}{2014}]{Gourgouliatos2014}
Gourgouliatos K.~N.,  Cumming A.,  2014, \mn@doi [MNRAS]
  {10.1093/mnras/stu2140}, 446, 1121

\bibitem[\protect\citeauthoryear{Gourgouliatos, Cumming, Reisenegger, Armaza,
  Lyutikov  \& Valdivia}{Gourgouliatos et~al.}{2013}]{Gourgouliatos2013}
Gourgouliatos K.~N.,  Cumming A.,  Reisenegger A.,  Armaza C.,  Lyutikov M.,
  Valdivia J.~A.,  2013, \mn@doi [MNRAS] {10.1093/mnras/stt1195}, 434, 2480

\bibitem[\protect\citeauthoryear{Gourgouliatos, Wood  \&
  Hollerbach}{Gourgouliatos et~al.}{2016}]{Gourgouliatos2016}
Gourgouliatos K.~N.,  Wood T.~S.,   Hollerbach R.,  2016, \mn@doi [PNAS]
  {10.1073/pnas.1522363113}, 113, 1522363113

\bibitem[\protect\citeauthoryear{Graber, Andersson, Glampedakis  \&
  Lander}{Graber et~al.}{2015}]{Graber2015}
Graber V.,  Andersson N.,  Glampedakis K.,   Lander S.~K.,  2015, \mn@doi
  [MNRAS] {10.1093/mnras/stv1648}, 453, 671

\bibitem[\protect\citeauthoryear{Grad \& Rubin}{Grad \& Rubin}{1958}]{Grad1958}
Grad H.,  Rubin H.,  1958, \mn@doi [Journal of Nuclear Energy]
  {10.1016/0891-3919(58)90139-6}, 7, 284

\bibitem[\protect\citeauthoryear{Gusakov}{Gusakov}{2019}]{Gusakov2019}
Gusakov M.~E.,  2019, \mn@doi [MNRAS] {10.1093/mnras/stz657}, 485, 4936

\bibitem[\protect\citeauthoryear{Gusakov \& Dommes}{Gusakov \&
  Dommes}{2016}]{Gusakov2016}
Gusakov M.~E.,  Dommes V.~A.,  2016, \mn@doi [PhysRevD]
  {10.1103/PhysRevD.94.083006}, 94, 083006

\bibitem[\protect\citeauthoryear{Gusakov, Kantor  \& Ofengeim}{Gusakov
  et~al.}{2017}]{Gusakov2017}
Gusakov M.~E.,  Kantor E.~M.,   Ofengeim D.~D.,  2017, \mn@doi [PhysRevD]
  {10.1103/PhysRevD.96.103012}, 96, 103012

\bibitem[\protect\citeauthoryear{Haensel}{Haensel}{1995}]{Haensel1995}
Haensel P.,  1995, \mn@doi [SSRv] {10.1007/BF00751429}, 74, 427

\bibitem[\protect\citeauthoryear{Hollerbach \& R{\"{u}}diger}{Hollerbach \&
  R{\"{u}}diger}{2002}]{Hollerbach2002}
Hollerbach R.,  R{\"{u}}diger G.,  2002, \mn@doi [MNRAS]
  {10.1046/j.1365-8711.2002.05905.x}, 337, 216

\bibitem[\protect\citeauthoryear{Hoyos, Reisenegger  \& Valdivia}{Hoyos
  et~al.}{2008}]{Hoyos2008}
Hoyos J.,  Reisenegger A.,   Valdivia J.~A.,  2008, \mn@doi [A{\&}A]
  {10.1051/0004-6361:200809466}, 487, 789

\bibitem[\protect\citeauthoryear{Hoyos, Reisenegger  \& Valdivia}{Hoyos
  et~al.}{2010}]{Hoyos2010}
Hoyos J.,  Reisenegger A.,   Valdivia J.~A.,  2010, \mn@doi [MNRAS]
  {10.1111/j.1365-2966.2010.17237.x}, 408, 1730

\bibitem[\protect\citeauthoryear{Kantor \& Gusakov}{Kantor \&
  Gusakov}{2018}]{Kantor2017}
Kantor E.~M.,  Gusakov M.~E.,  2018, \mn@doi [MNRAS] {10.1093/mnras/stx2682},
  473, 4272

\bibitem[\protect\citeauthoryear{Kaspi}{Kaspi}{2010}]{Kaspi2010}
Kaspi V.~M.,  2010, \mn@doi [PNAS] {10.1073/pnas.1000812107}, 107, 7147

\bibitem[\protect\citeauthoryear{Lander \& Gourgouliatos}{Lander \&
  Gourgouliatos}{2019}]{Lander2019}
Lander S.~K.,  Gourgouliatos K.~N.,  2019, \mn@doi [MNRAS]
  {10.1093/mnras/stz1042}, 486, 4130

\bibitem[\protect\citeauthoryear{Lander \& Jones}{Lander \&
  Jones}{2009}]{Lander2009}
Lander S.~K.,  Jones D.~I.,  2009, \mn@doi [MNRAS]
  {10.1111/j.1365-2966.2009.14667.x}, 395, 2162

\bibitem[\protect\citeauthoryear{Lander \& Jones}{Lander \&
  Jones}{2012}]{Lander2012}
Lander S.~K.,  Jones D.~I.,  2012, \mn@doi [MNRAS]
  {10.1111/j.1365-2966.2012.21213.x}, 424, 482

\bibitem[\protect\citeauthoryear{Marchant, Reisenegger  \&
  Akg{\"{u}}n}{Marchant et~al.}{2011}]{Marchant2011}
Marchant P.,  Reisenegger A.,   Akg{\"{u}}n T.,  2011, \mn@doi [MNRAS]
  {10.1111/j.1365-2966.2011.18874.x}, 415, 2426

\bibitem[\protect\citeauthoryear{Marchant, Reisenegger, Valdivia  \&
  Hoyos}{Marchant et~al.}{2014}]{Marchant2014}
Marchant P.,  Reisenegger A.,  Valdivia J.~A.,   Hoyos J.~H.,  2014, \mn@doi
  [ApJ] {10.1088/0004-637X/796/2/94}, 796, 94

\bibitem[\protect\citeauthoryear{Markey \& Tayler}{Markey \&
  Tayler}{1973}]{Markey1973}
Markey P.,  Tayler R.~J.,  1973, \mn@doi [MNRAS] {10.1093/mnras/163.1.77}, 163,
  77

\bibitem[\protect\citeauthoryear{Mitchell, Braithwaite, Reisenegger, Spruit,
  Valdivia  \& Langer}{Mitchell et~al.}{2015}]{Mitchell2015}
Mitchell J.~P.,  Braithwaite J.,  Reisenegger A.,  Spruit H.,  Valdivia J.~A.,
   Langer N.,  2015, \mn@doi [MNRAS] {10.1093/mnras/stu2514}, 447, 1213

\bibitem[\protect\citeauthoryear{Ofengeim \& Gusakov}{Ofengeim \&
  Gusakov}{2018}]{Ofengeim2018}
Ofengeim D.~D.,  Gusakov M.~E.,  2018, \mn@doi [PhysRevD]
  {10.1103/PhysRevD.98.043007}, 98, 043007

\bibitem[\protect\citeauthoryear{Passamonti, Akg{\"{u}}n, Pons  \&
  Miralles}{Passamonti et~al.}{2017a}]{Passamonti2017}
Passamonti A.,  Akg{\"{u}}n T.,  Pons J.~A.,   Miralles J.~A.,  2017a, \mn@doi
  [MNRAS] {10.1093/mnras/stw2936}, 465, 3416

\bibitem[\protect\citeauthoryear{Passamonti, Akg{\"{u}}n, Pons  \&
  Miralles}{Passamonti et~al.}{2017b}]{Passamonti2017b}
Passamonti A.,  Akg{\"{u}}n T.,  Pons J.~A.,   Miralles J.~A.,  2017b, \mn@doi
  [MNRAS] {10.1093/mnras/stx1192}, 469, 4979

\bibitem[\protect\citeauthoryear{Pethick}{Pethick}{1992}]{Pethick1992}
Pethick C.,  1992, in Structure and Evolution of Neutron Stars. p.~115

\bibitem[\protect\citeauthoryear{Pons \& Geppert}{Pons \&
  Geppert}{2007}]{Pons2007}
Pons J.~A.,  Geppert U.,  2007, \mn@doi [A{\&}A] {10.1051/0004-6361:20077456},
  470, 303

\bibitem[\protect\citeauthoryear{Pons, Miralles  \& Geppert}{Pons
  et~al.}{2009}]{Pons2009}
Pons J.~A.,  Miralles J.~A.,   Geppert U.,  2009, \mn@doi [A{\&}A]
  {10.1051/0004-6361:200811229}, 496, 207

\bibitem[\protect\citeauthoryear{Reisenegger}{Reisenegger}{1995}]{Reisenegger1995}
Reisenegger A.,  1995, \mn@doi [ApJ] {10.1086/175480}, 442, 749

\bibitem[\protect\citeauthoryear{Reisenegger}{Reisenegger}{2007}]{Reisenegger2007}
Reisenegger A.,  2007, \mn@doi [AN] {10.1002/asna.200710848}, 328, 1173

\bibitem[\protect\citeauthoryear{Reisenegger}{Reisenegger}{2009}]{Reisenegger2009}
Reisenegger A.,  2009, \mn@doi [A{\&}A] {10.1051/0004-6361/200810895}, 499, 557

\bibitem[\protect\citeauthoryear{Reisenegger \& Goldreich}{Reisenegger \&
  Goldreich}{1992}]{Reisenegger1992}
Reisenegger A.,  Goldreich P.,  1992, \mn@doi [ApJ] {10.1086/171645}, 395, 240

\bibitem[\protect\citeauthoryear{Shafranov}{Shafranov}{1966}]{Shafranov1966}
Shafranov V.~D.,  1966, RvPP, 2, 103

\bibitem[\protect\citeauthoryear{Tauris \& van~den Heuvel}{Tauris \& van~den
  Heuvel}{2006}]{Tauris2006}
Tauris T.~M.,  van~den Heuvel E. P.~J.,  2006, in Lewin W.,  van~der Klis M.,
  eds, , Compact stellar X-ray sources.
pp 623--665

\bibitem[\protect\citeauthoryear{Tayler}{Tayler}{1973}]{Tayler1973}
Tayler R.~J.,  1973, \mn@doi [MNRAS] {10.1093/mnras/161.4.365}, 161, 365

\bibitem[\protect\citeauthoryear{Thompson \& Duncan}{Thompson \&
  Duncan}{1995}]{Thompson1995}
Thompson C.,  Duncan R.~C.,  1995, MNRAS, 275, 255

\bibitem[\protect\citeauthoryear{Thompson \& Duncan}{Thompson \&
  Duncan}{1996}]{Thompson1996}
Thompson C.,  Duncan R.~C.,  1996, \mn@doi [ApJ] {10.1086/178147}, 473, 322

\bibitem[\protect\citeauthoryear{Vigan{\`{o}}, Pons  \& Miralles}{Vigan{\`{o}}
  et~al.}{2012}]{Vigano2012}
Vigan{\`{o}} D.,  Pons J.~A.,   Miralles J.~A.,  2012, \mn@doi [CPC]
  {10.1016/j.cpc.2012.04.029}, 183, 2042

\bibitem[\protect\citeauthoryear{Vigan{\`{o}}, Rea, Pons, Perna, Aguilera  \&
  Miralles}{Vigan{\`{o}} et~al.}{2013}]{Vigano2013}
Vigan{\`{o}} D.,  Rea N.,  Pons J.~A.,  Perna R.,  Aguilera D.~N.,   Miralles
  J.~A.,  2013, \mn@doi [MNRAS] {10.1093/mnras/stt1008}, 434, 123

\bibitem[\protect\citeauthoryear{Wright}{Wright}{1973}]{Wright1973}
Wright G. A.~E.,  1973, \mn@doi [MNRAS] {10.1093/mnras/162.4.339}, 162, 339

\bibitem[\protect\citeauthoryear{Yakovlev \& Shalybkov}{Yakovlev \&
  Shalybkov}{1990}]{Yakovlev1990}
Yakovlev D.~G.,  Shalybkov D.~A.,  1990, SvA Lett., 16, 86

\makeatother
\end{thebibliography}





\bsp	
\label{lastpage}
\end{document}